# Portable Low-cost MRI System based on Permanent Magnets/Magnet Arrays

Shao Ying HUANG[1,2], Zhi Hua REN[1], Sergei OBRUCHKOV[3], Jia GONG[1], Robin DYKSTRA[3], Wenwei YU[4]

[1] Singapore University of Technology and Design, Singapore

[2] Department of Surgery, National University of Singapore, Singapore

[3] Victoria University of Wellington, New Zealand

[4] Chiba University, Japan

**Abstract:**

Portable low-cost magnetic resonance imaging (MRI) systems have the potential to enable "point-of-care" and timely MRI diagnosis, and to make this imaging modality available to routine scans and to people in underdeveloped countries and areas. With simplicity, no maintenance, no power consumption, and low cost, permanent magnets/magnet arrays/magnet assemblies are attractive to be used as a source of static magnetic field to realize the portability and to lower the cost for an MRI scanner. However, when taking the canonical Fourier imaging approach and using linear gradient fields, homogeneous fields are required in a scanner, resulting in the facts that either a bulky magnet/magnet array is needed, or the imaging volume is too small to image an organ if the magnet/magnet array is scaled down to a portable size. Recently, with the progress on image reconstruction based on non-linear gradient field, static field patterns without spatial linearity can be used as spatial encoding magnetic fields (SEMs) to encode MRI signals for imaging. As a result, the requirements for the homogeneity of the static field can be relaxed, which allows permanent magnets/magnet arrays with reduced sizes, reduced weight to image a bigger volume covering organs such as a head. It offers opportunities of constructing a truly portable low-cost MRI scanner. For this exciting potential application, permanent magnets/magnet arrays have attracted increased attention recently. A magnet/magnet array is strongly associated with the imaging volume of an MRI scanner, image reconstruction methods, and RF excitation and RF coils, etc. through field patterns and field homogeneity. This paper offers a review of permanent magnets and magnet arrays of different kinds, especially those that can be used for spatial encoding towards the development of a portable and low-cost MRI system. It is aimed to familiarize the readers with relevant knowledge, literature, and the latest updates of the development on permanent magnets and magnet arrays for MRI. Perspectives on and challenges of using a permanent magnet/magnet array to supply a patterned static magnetic field, which does not have spatial linearity nor high field homogeneity, for image reconstruction in a portable setup are discussed.

## I. Introduction

Magnetic resonance imaging (MRI) shows advantages of good soft tissue contrast, non-ionizing radiation, capability to image at an arbitrary plane, etc. It is a unique medical imaging modality that is important to diagnoses and treatments of a wide range of diseases, which other imaging modalities, such as ultra sound and computed tomography (CT) cannot offer with a similar level of tissue contrast and/or a similar level of safety.

A MRI human scanner commonly seen in a hospital has a superconducting magnet to supply the main magnetic field, $\boldsymbol{B_0}$. In a superconducting magnet, there is a long cylindrical bore which is used for housing human body for imaging. Fig. 1 shows an example of such a scanner (Siemens 3T MAGNETOM Prisma). A superconducting-magnet-based MRI scanner takes multiple rooms to sit the whole system, and it is costly to purchase and maintain with the purchase price approaching $1 million U.S. dollars per tesla (T) of magnetic field.

In a conventional MRI scanner, because it is bulky, normally with a weight of greater than 10 tons, it cannot support in-time and on-site medical imaging, for example, in the situations when space, electric power, and other resources are limited such as imaging the wounded after a natural disaster or imaging casualties in a military field hospital, or imaging at remote locations or in rural areas. Because of the bulkiness, current MRI scanners cannot provide imaging in Intensive Care Units (ICU) where there are dangers associated with transporting patients in the units. Moreover, due to a high cost, it is not affordable by every patient who needs it, neither by those who need it regularly, e.g. the elderly who need regular and repeated scans. Because that an MRI scanner is expensive, it is normally not available in a community hospital or a clinic in a village. A portable MRI scanner (trolley-portable or man-portable) with a reasonable imaging volume for scanning human and at a relatively low cost will be of importance to make MRI available in more situations that medical imaging is needed and to allow a bigger population to enjoy the benefits brought by this imaging modality.

Efforts have been made to make a traditional MRI scan more flexible and to reduce the size of an MRI scanner. A long-bore scanner shows limitations in clinical practise. It cannot support scanning of patients who have claustrophobia, that of those who have non-MRI compatible implants, or that of obese patients. It does not support interventional medical procedure either. Open whole-body MRI scanners address the aforementioned problems. For an open MRI, different types of magnets are used, for example, superconducting magnet (0.6 T – 1.2 T, at a scale of 10 tons) [1], electromagnet (with resistive coils, 0.3 T - 0.6 T) [2], or permanent magnet ( $\leq$ 0.35T, at a scale of 15 tons) [3]. Fig. 2 shows the examples of open MRI, such as a short-bore MRI [4] (Fig. 2 (a), superconducting-magnet-based), C-shaped MRI [5] (Fig. 2 (c), permanent-magnet-based), the MRI scanners where the subject can sit up right (Fig. 2 (c) [6], electromagnet-based, and (d) [7], superconducting-magnet-based). Open MRI scanners offer flexibilities for scanning and the sizes are reduced from a long-bore scanner. However, they are still of room size (at least a room of 30 $m^2$ is needed), heavy, and expensive.

The bulkiness of a conventional MRI scanner (a long-bore or open MRI) is mainly because of the magnets that supplies the $\boldsymbol{B_0}$-field in a reasonable imaging volume for a human scan. $\boldsymbol{B_0}$-field is required to be homogeneous so that linear gradient fields can be applied by gradient coils to encode MRI signals for image reconstruction using Fourier imaging method [8]. A high field strength is desired because it leads to a high signal-to-noise ratio (SNR) and a high spatial resolution. The range of gradient fields is controlled to be small which corresponds to a narrow frequency bandwidth and a high quality-factor (Q-factor) of the radiofrequency (RF) system, including the RF excitation, RF transmit and receive coils, and RF reception. In a conventional MRI system, when Fourier imaging method is used, the $\boldsymbol{B_0}$-field provided by a magnet is required to be less than a few parts per million (ppm) in the imaging volume, normally of a size of 40 x 40 x 40 cm to 50 x 50 x 50 cm for a decent whole-body scan. A high homogeneity of $\boldsymbol{B_0}$ in a body-size volume is normally obtained at a price of a large and heavy magnet (either superconducting magnet, electromagnet, or permanent magnet) where a lot of shimming structures are needed.

Efforts have been made to further reduce the size of an MRI scanner. A magnet is the biggest component in an MRI system. The key to make a scanner lighter and smaller is to have the magnet and the corresponding hardware (gradient coils, RF coils, and detection coils) built around the organ under scanning. For example, for head imaging, the apparatus is built around the head. O-scan from Esaote [9] as shown in Fig. 3 is an example. A permanent magnet (precisely speaking it is a permanent magnet array/assembly which consists of multiple magnets) are used (field strength: 0.31T, Gradient: ± 20 mT/m) to support Fourier image reconstruction. O-scan targets on arms and legs, therefore, the size of the system is dramatically reduced. It requires an installation space of 9 $m^2$ and it weighs 1.4 tons. A similar approach was proposed by using a superconducting magnet. In 1995, S. Pissanetzky from University of Houston Clear Lake proposed to build a magnet around the organ under imaging [10]. A bagel-shaped magnet, with an outer diameter of 1.1 m, an inner diameter of 20 cm, a thickness of 15 cm, and a weight of 0.5 ton, was proposed for breast imaging in a field of view (FoV) of 10 cm in diameter with a field strength of 1 T and an inhomogeneity of less than 10 ppm. Fig. 4 (a) shows the original illustration in the patent. Fig. 4 (b) shows a helmet-shaped magnet proposed in the same patent [10][5] for head imaging. This magnet has an outer diameter of 0.6 m, an inner diameter of 20 cm, a thickness of 35 cm, and a weight of 0.5 ton. It may generate a field strength of 2 T and an inhomogeneity of less than 10 ppm in a FoV of 20 cm in diameter. Recently, J. T. Vaughan and his team proposed a liquid nitrogen-cooled magnet with a size of 36 cm x 71 cm with a weight of 250 kg for head imaging [11]. The field strength is 1.5 T with 50 ppm in 10 cm diameter of spherical volume (DSV). It does not require liquid helium and can operate "off-the-grid". The size of the magnet is reduced by half compared to the magnet proposed by Pissanetzky although the FoV is much smaller and the inhomogeneity is much higher. When the FoV increases to be of a head size, the inhomogeneity increases. The high inhomogeneity may impose difficulties to the design of RF coils, and challenge the traditional Fourier image reconstruction.

Building the magnet and the corresponding hardware around the organ under imaging does effectively reduce the size of the system from tens of tons to less than 250 kg (0.25 ton) by using either permanent magnets or superconducting magnets, when traditional Fourier imaging is applied and the homogeneity of the $\boldsymbol{B_0}$-field has to be guaranteed. Comparing permanent magnets to super conducting magnets, no cooling system is needed and there is no power consumption. Meanwhile, it is less expensive. Therefore, a permanent magnet can be a good candidate to lower the cost of an MRI scanner. However, a permanent magnet is still large and heavy when the field homogeneity have to be guaranteed for Fourier imaging (using linear gradient field and Fourier transfer). A scanner using a permanent magnet to have a strength of 0.35T with homogeneity in a FoV of $40\times 40 \times 40$ cm has a weight of 17.6 tons[3]. When the FoV is reduced to 14 cm for a field with homogeneity for arms or legs, as shown in the O-scan by Esaote in Fig. 3, the weight of the magnet can be reduced to 1.4 tons, which is still heavy to move around. On the other hand, when a permanent magnet (C-shaped array) is scaled down to a table-top size, the FoV has to shrink to a volume of 1.27 x 1.27 x 1.90 $cm^3$ [12] which is not practical to image any organ. Comparing permanent magnets to electromagnets, they are both low-cost. Different from permanent magnets that have no power consumption and does not need a cooling system, for an electromagnet, it generates static magnetic fields using DC currents with a relatively simple cooling system. Although it could be installed relatively easily and can be switched off anytime, an electromagnet provides relatively weak field strength compared to a permanent magnet. The reason is that it dissipates considerable amount of heat, which limits its input power and consequently capping the field strength it can generate. Besides a limited field strength, the size of an electromagnet has to be much larger than the FoV to maintain a certain field homogeneity [13, 14]. In [13], a custom built biplanar electromagnet [14] has a cross-sectional outer diameter of 220 cm, a width of 90 cm, and a weight of about 340 kg for generating an average field strength of 6.5 mT and a homogeneity of about 100 ppm within a FOV with a diameter of 40 cm (head-sized).

With no cooling systems, no power consumption, and low cost, permanent magnets/permanent magnet arrays are attractive to be used as the source of static magnetic field for MRI imaging. However, for decades, with a traditional setup and imaging method where homogeneous $\boldsymbol{B_0}$-field and spatially linear gradient fields are used, permanent magnets that supply $\boldsymbol{B_0}$-fields are inevitably large to achieve a field homogeneity that is needed in a FoV for body imaging. Consequently, they cannot offer a portable solution for imaging human. Recently, MRI image reconstructions based on non-linear gradient field have been proposed to overcome physiological limitation of the conventional spatial linear gradient setup, e.g. to reduce peripheral nerve stimulation. Examples of the approaches based on a non-linear gradient field are the parallel imaging technique using localized gradients (PatLoc) imaging [15] , image encoding using arbitrary shaped, curvilinear, and nonbijective magnetic fields [16] or using multipolar fields for radial imaging [17], and O-space imaging [18, 19]. In these approaches, static field patterns without spatial linearity are used to encode MRI signals for imaging. Such fields are named spatial encoding magnetic fields (SEMs) [15]. With the progress on imaging reconstruction based on non-linear gradient fields, the requirements on the $\boldsymbol{B_0}$-field homogeneity can be relaxed when $\boldsymbol{B_0}$ is seen as the sum of a homogeneous field and a gradient field and used as SEMs. Consequently, if a permanent magnet/magnet array is used to provide such a $\boldsymbol{B_0}$-field, the SEM, its size and weight can be reduced and meanwhile the imaging volume can be increased, offering opportunities to construct a truly portable low-cost MRI scanner for human scanning. In [20], a long Halbach-type magnet array was designed to provide a SEM for head imaging. In [21] and [22], a short Halbach-type array and an Aubert ring-pair aggregate were proposed to provide a SEM for two-dimensional (2D) head imaging, respectively.

In an MRI scanner with expected portability and a low cost when a permanent magnet/magnet array is used to provide a spatially non-linear SEM, the magnet/magnet array determines the field strength, the field pattern, i.e. distributions of fields, and the field homogeneity. Both the field strength and the field pattern determine the SNR and the resolution of the system [16]. For a specific field pattern, image reconstruction algorithms need to be tailored for imaging. For the field homogeneity, because the field is less homogeneous, the working bandwidth of the control circuit and that of the RF coils increase. As a result, the RF hardware has to be re-engineered to obtain a corresponding wider working bandwidth. In a portable setup for MRI imaging, the magnet/magnet array decides the quality of the image, and heavily influences the designs of the rest of the MRI system, playing a key role.

In this paper, permanent magnets and permanent magnet arrays with different configurations are reviewed. It is aimed to familiarize the readers with relevant knowledge, literature, and the latest updates of the development on permanent magnets/magnet arrays for MRI, especially for a portable low-cost MRI. The challenges of using a permanent magnet/magnet array to supply static magnetic fields as SEMs with non-linear gradient for image reconstruction in a portable setup are discussed.

## II. Materials and Methods

### II.a Permanent Magnets

A permanent magnet array consists of permanent magnets with designed shapes, polarizations, and a specific configuration. A permanent magnet is an artificial object made from a material that is magnetized and creates its own persistent magnetic field even in the absence of an applied magnetic field. It is made of ferromagnetic or ferrimagnetic materials that show spontaneous magnetizations without an external magnetic field below Curie temperature. Examples of such kind of materials are iron, nickel, cobalt, and alloys of rare-earth metals.

There are different types of permanent magnets. One type is **naturally occurring ferromagnets**, such as magnetite (or lodestone), nickel, cobalt, and rare-early metals such as gadolinium and dysprosium (at a very low temperature due to their low Curie temperatures). They are used in the early experiments with magnetism. With the advancement of technology, composites based on the natural magnetic materials are produced, with improved magnetic field strength and mechanical properties. They are the second type of permanent magnets. The magnet field strength of this type of magnets can reach 1 T. One example is **ceramic magnets** that are made of sintered composite of powdered iron oxide and barium/strontium carbonate ceramic. They are inexpensive and can be easily mass produced. They are non-corroding but brittle. There are other examples such as **injection-moulded magnets** which are composite of various types of resin and magnetic powders, **alnico magnets** that are made by casting or sintering a combination of aluminium, nickel, and cobalt with iron and small amounts of other elements. Another example is **flexible magnets** that are composed of a high-coercivity ferromagnetic compound mixed with a plastic binder.

The third type is **rare-earth magnets**. Rare-earth magnets are the strongest type of permanent magnets made from alloy of rare-earth elements that are the fifteen metallic chemical elements with atomic numbers from 57 to 71 in the periodic table. They were proposed at the beginning of 1980s. Their magnetic field can exceed 1 T. The high magnetic field comes from the rare-earth elements (e.g. Scandium, ${}_{21}Sc$, Yttrium, ${}_{39}Y$) that have atoms that retain high magnetic moments in the solid state, which is a consequence of incomplete filling of the f-shell allowing up to seven unpaired electrons with aligned spins. The rare-earth elements show low Curie temperature above which the material losses magnetism. However, when they form compounds with transition metals (e.g. iron, nickel, and cobalt), the Curie temperatures of the compounds increases and becomes higher than the room temperature. There are mainly two types of rare-earth magnets, **Samarium (SmCo)** and **neodymium (NdFeB)**. NdFeB permanent magnets were invented after SmCo. They have higher energy products compared to SmCo [23]. However, NdFeB is much more sensitive to the change in temperature than SmCo. NdFeB has a temperature coefficient of $-1.1 \times 10^{-3}$/°C whereas SmCo has one of $-0.3 \times 10^{-3}$/°C (Temperature coefficient is a parameter to quantify the temperature sensitivity of a magnetic material to a temperature variation. It is defined as $(\Delta B/B)/\Delta T$ where $\Delta B$ is the change in the magnetic field, $B$ is magnetic field, and $\Delta T$ is the change in temperature). Table I below shows their magnetic properties where remanence ($B_r$) measures the strength of the magnetic field, coercivity ($H_c$) is the resistance to becoming demagnetized of the material, energy product, $(BH)_{max}$, is the density of magnetic energy; and $T_c$ is Curie temperature.

**Table I. Magnetic Properties of Rare-earth Magnets**

| Magnet | $B_r$ (T) | $H_c$ (kA/m) | $(BH)_{max}$ (kJ/m$^3$) | $T_c$ (°C) |
|---|---|---|---|---|
| **SmCo (sintered)** | 0.8–1.1 | 600–2000 | 120–200 | 720 |
| **NdFeB (sintered)** | 1.0–1.4 | 750–2000 | 200–440 | 70–200 |
| **NdFeB (bonded)** | 0.6–0.7 | 600–1200 | 60–100 | 80–150 |

Sintered NdFeB has a Curie temperature ranging from 70°C to 200°C. Depending on the Curie temperature, it is grouped into N35-N52, N33M-N48M, N30H-N45H, N30SH-N42SH, N30UH-N35UH, N28EH-N35EH which corresponds to the Curie temperature of 70°C - 80°C, 100°C, 120°C, 150°C, 180°C, and 200°C, respectively. The magnets are more brittle for the group that have a higher Curie temperature. Within each category, they are further graded based on the magnetic field strength. The N35-N52 group is the most commonly commercially available.

Bonded NdFeB lies between sintered NdFeB and ferrite in properties and is more stable. It is free from further machining and easy for pressing into different shapes such as rings, arcs, etc. It has a much higher corrosion resistance in comparison with sintered NdFeB. Bonded magnets offer less flux than sintered magnets but can be net-shape formed into intricately shaped parts and do not suffer significant eddy current losses. Compared to NdFeB, SmCo is weaker but more suitable to work in high temperature. It is anti-corrosive.

In a magnet array, rare-earth magnets are popular to be used because of the relatively high magnetic field strength they provide among different types of permanent magnets. Due to the availability, NdFeB, especially the N35-N52 group of sintered NdFeB, is a popular option.

### II.b Permanent Magnet Arrays for MR Imaging with Homogeneous $B_0$

When traditional Fourier MR imaging methods with gradient fields are used for imaging, the main field $B_0$ are required to be homogeneous. Permanent magnet/magnet arrays that supply homogeneous dipolar fields, such as the **C-shaped or H-shaped dipolar magnet** and a **Halbach array**, have been used for MRI imaging. They are reviewed as follows.

#### b1. Dipolar Permanent Magnet/Magnet Array (C-shaped/H-shaped Magnet)

A dipolar permanent magnet/magnet array, such as a C-shaped or a H-shaped magnet, is a magnetic circuit that consists of both iron and permanent magnets where the iron is used to guide the magnetic flux. It has two poles and they generate homogeneous magnetic fields between the poles. The magnetic field between the poles is used to supply the main magnetic field for MR imaging. Fig. 5 (a) and (c) show the 3D views of a C-shaped and an H-shaped magnet, respectively. Fig. 5 (b) and (d) shows their side views where magnetic fluxes inside the yoke and between the poles are indicated. The field can be vertical, as shown in Fig. 5, or horizontal. It is used in commercial MRI scanners which belongs to the category of open MRI. An example of a commercial product is shown in Fig. 2 (b) where the field supplied by the magnet is vertical.

A C-shaped magnet with a small pole size as shown in Fig. 6 is a conventional magnet. It is widely used in NMR measurements [24 - 26]. In [12], a C-shaped permanent magnet, which is an array, with an increased pole size scalable to a full body MRI was proposed. The technical details were reported using a scaled down table-top size array. For the reported table-top C-shaped magnet array, it has a size of 16” x 20” x 7”, providing a magnetic field of 0.21 T with a homogeneity of 20 ppm within a cylindrical region of 0.5” in diameter and 0.75” in length. Fig. 7 (a) and (b) show a photo and the cross-sectional view (with dimensions) of the miniaturized C-shaped magnet, respectively. As shown in Fig. 7 (b), the C-shaped magnet consists of a rectangular C-arm connecting two iron necks, two NdFeB poles, and two pole faces with an air gap in between. The size of the air gap is 7” × 4”. Both the poles and the necks are cylindrical, the pole has a diameter of 7” and the necks are tapered from a diameter of 7” to one of 4”. The tapered diameter of the neck at the junction to the C-arm is for a reduction in iron volume so as to reduce the overall weight of the system. Both the dimensions of the neck and the C-arm are optimized so that the iron does not saturate, and the reluctance does not increase dramatically. Pole faces are designed to re-focus magnetic field lines toward the C-gap.

In [12], the magnet poles were constructed using 330 NdFeB blocks with a dimension of 1” × 1” × 0.5”. Each block was numbered, and the energy was measured. The magnets were stacked in groups of three in order to minimize energy variation. This resulted in several groups of magnetization energy which were placed symmetrically around the pole pieces. The pole pieces were designed using numerical solutions, the 2D Pandira code from Los Alamos

National Lab. Shimming was achieved by adjustment of the pole pieces and by using four electrical shim coils. Based on the construction of a C-shaped magnet, strictly speaking, it is a magnet array or assembly rather than a single magnet. Using the proposed C-shaped magnet array in [12], a table-top MRI system was reported in [27] with the mentioned imaging volume (a cylindrical volume with a diameter of 0.5“ and a length of 0.75”). A C-shaped permanent magnet was applied to image plants [33] and arms [34] in Terada Lab in the University of Tsukuba. Fig. 8 (a) shows the magnet for scanning a living tree and Fig. 8 (b) shows the whole MRI scanning system. The magnet (NEOMAX Co., Osaka, Japan) has a magnetic field of 0.21 T with a homogeneity of 34.6 ppm in a volume of 20 cm x 20 cm x 12 cm diameter ellipsoidal volume and a gap size of 160 mm. The size of the magnet is 70 cm x 50 cm x 45 cm, and the weight is 520 kg. Fig. 8 (c) shows the C-shaped magnet for scanning a hand. This magnet has an imaging volume of 12 cm x 16 cm x 5 cm and a field strength of 0.3 T with a homogenous of 16 ppm. The size of the magnet is 52 cm x 62 cm x 52 cm and the weight is around 450 kg. Comparing these two C-shaped magnets, an increase in magnetic field is obtained by adding magnetic materials at a price of increasing the size and the weight of the magnet.

For an H-shaped magnet array, a scaled down table-top prototype was built by the Institute of Electrical Engineering of the Chinese Academy of Sciences in Beijing [35]. Fig. 9 (a) and (b) shows the photo and the cross-sectional view of the magnet, respectively. The field strength is 0.19 T and the weight is 13 kg. As shown in Fig. 9 (b), there are two iron pole faces, two rare earth magnets, and an iron case forming a magnetic circuit with a designed flow of the magnetic flux. The image volume is between the two pole faces. The distance between the pole faces is 4 cm, and the homogeneity is about 50 ppm over a DSV of 1 cm. It was used in a table-top MRI scanner developed in Martinos Center (Massachusetts General Hospital) for Biomedical Imaging [36]. Compared to a C-shaped magnet, the H-shaped has two paths to guide the flux in the iron yoke. The struct ion is symmetric, and it is mechanically stable.

**b2. Halbach Array**

A Halbach permanent magnet array [37 - 39] is a special arrangement of permanent magnets in which the array has a strong magnetic field pattern on one side of the array while a minimized field on the other side. It can be one-dimensional (1D) (linear Halbach array), 2D (Halbach cylinder), and 3D (Halbach sphere). A linear Halbach array was first reported by J. C. Mallinson in [34] in 1973 while independently reported by K. Halbach in [32, 33]. In [34], it was described by Mallinson as "one-sided flux" that was discovered with "curiosity". He recognized from this discovery the potential for significant improvements in magnetic tape technology. On the other hand, with the new development of magnetic materials such as NdFeB and SmCo at the beginning of 1980s [23], K. Halbach saw the opportunities of creating strong magnetic field using the rare-earth magnets with high energy products. Halbach’s invention is mainly for focusing particle accelerator beams. It is more well-known thus this type of magnet array is named after him. Fig. 10 (a), (b), and (c) show a 1D, 2D, and 3D Halbach array, respectively. The Halbach array in Fig. 10 (b) and (c) are dipolar.

For the application of NMR and MRI, Halbach cylinder is widely used [20, 21, 35, 36] . An ideal Halbach cylinder has magnetization, $\vec{M}$, that is continuously changed along the $\phi$-direction in the cylindrical coordinate system. Fig. 11 (a) shows the relation of the magnetization in the cylinder with the angles. The magnetization distribution in a cylinder is expressed as follows [37]. At a location on the cylinder with an angle of $\phi$ with respect to the x-axis where $\emptyset = 0^o$, the magnetization is expressed as,

$$\vec{M} = M_r\hat{r} + M_\emptyset\hat{\emptyset} \quad (1)$$

where $M_r = |\vec{M}|\cos(\emptyset_m)$ , $M_\emptyset = \pm|\vec{M}|\sin(\emptyset_m)$ , are the *r*- and ϕ-components of the magnetization, $\emptyset_m$ is the angle of the direction of the magnetization with respect to the angle $\emptyset$, the positive sign, '+', corresponds to an internal-field Halbach cylinder with zero field outside the cylinder (a dipolar case is shown in Fig. 11 (b)), and the negative sign, '-', corresponds to an external-field Halbach cylinder with zero field inside the cylinder (a dipolar case is shown in Fig. 11 (c)). The relation of $\emptyset$ and $\emptyset_m$ is the following,

$$\emptyset_m = (1 \pm n)\emptyset \qquad (2)$$

where n is the number of pole-pairs or modes, $n = \pm 1$ corresponds to a dipolar Halbach array, $n = \pm 2$ corresponds to a quadrupolar Halbach array. Fig. 11 (b) shows the side view of an internal-field cylinder when n = 1 with a dipolar field pattern and Fig. 11 (d) shows that of an internal-field cylinder when n = 2 with a quadrupolar field pattern. In a dipolar internal-field cylinder as shown in Fig. 11 (b), the dipolar internal field is homogeneous [32, 33, 38]. When the cylinder is infinitely long, the magnetic field in the air tunnel is expressed as below [38],

$$\vec{B} = \hat{x}B_x \qquad (3)$$

The magnetic field is along the $\emptyset = 0$ direction (the x-axis as shown in Fig. 11 (a)) which is on the transversal plane of the cylinder.

$$B_x = B_r\, ln(\frac{r_{\text{out}}}{r_{\text{in}}}) \qquad (4)$$

where $B_r$ is the remanence of the magnetic materials, $r_{\text{out}}$ and $r_{\text{in}}$ are the outer and inner radius of the cylinder. Theoretically, an arbitrary magnetic field can be obtained inside the cylinder. However, the field is capped by coercivity limits at 3 - 4 T [49]. Although based on (4), $B_x$ is greater than the remanence, $B_r$, it varies with the logarithm of the ration, ($r_{\text{out}}/r_{\text{in}}$), which leads to a slow increase in $B_x$ with an increase in ($r_{\text{out}}/r_{\text{in}}$), and results in an impractical size and mass of the cylinder when a high $B_x$ is needed.

In terms of the fabrication of a Halbach cylinder, there are mainly two ways [37]. One is **the magnet ring approach** by moulding or sintering a complete ring magnet and subsequently impulse magnetising in a fixture, which produces a sinusoidally distributed magnetising field. The injection moulded NdFeB rings can be isotropic or anisotropic. The other way of implementation is **the discretized magnet approach** by using pre-magnetised magnet segments having the required magnetisation orientations to construct a cylinder. Three types of Halbach cylinders can be implemented, type (i) and (ii) are the results of taking the magnet ring approach, and type (iii) is the result by taking the discretized magnet approach. These three types are summarized below.

(i) Isotropic bonded NdFeB Halbach cylinders [40],

- Injection or compression moulded as ring magnet, and subsequently impulse magnetised in a sinusoidally distributed magnetising field
- Relatively easy to manufacture

(ii) Anisotropic bonded NdFeB Halbach cylinders [41]:

- Produced by orienting anisotropic NdFeB magnet powder during injection or compression moulding, and subsequently impulse magnetising in a Halbach distributed magnetising field
- The required aligning field for anisotropic NdFeB powder is < 1 T and the required magnetising field is > 3 T.

(iii) Segmented anisotropic Halbach cylinders [42, 43]

- Approximation to Halbach field distribution, dependent on the number of segments per pole
- A reduced field homogeneity
- High magnetic forces during assembly
- High resultant magnetic field

All three types are used for motors where the size of the cylinder is relatively small (diameter < 10 cm). In the discretized magnet approach, the cylinder is split into ring sectors for implementation [42]. Alternatively, it can be implemented using magnetic bars with a square cross section. A Halbach array implemented in this way is named "NMR Mandhala" (**M**agnet **A**rrangements for **N**ovel **D**iscrete **Ha**lbach **La**yout) [43]. For an "NMR Mandhala", the magnet array is sparse, and less mass is required. The sparsity and reduced mass of an "NMR Mandhala" comes at a price of a reduced field homogeneity of the array.

For a small volume, the magnet ring approach or the discretized magnet approach with a dense segmented cylinder is suitable to use. When the volume increases for applications for NMR and MRI, for example for imaging a human head, the discretized magnet approach with a sparse magnet array is more practical for the implementation for medical applications with a reasonable weight [20, 21].

In the discretized magnet approach, the cylinder is split into *M* sessions and at the $i^{th}$ session, a magnet with a magnetization of $\vec{M}^i$ is used. Fig. 11 (b) shows an illustration using a discretized Halbach cylinder with 12 sessions. The magnetization $\vec{M}^i$ is expressed as below,

$$\vec{M}^i = M_r^i\hat{r} + M_\emptyset^i\hat{\emptyset} \quad (5)$$

where $M_r^i = |\vec{M}^i|\cos(\emptyset_m^i)$, $M_r^i = \pm|\vec{M}^i|\sin(\emptyset_m^i)$, $\emptyset_m^i$ is the angle of the magnetization away from $\emptyset_i$ where $\emptyset_i$ is the angle of the center point at the $i^{th}$ session away from $\emptyset = 0^o$ along the x-axis. The relation of $\emptyset_m^i$ and $\emptyset_i$ are described as below, which depends on *n*, the number of pole-pairs or modes of the cylinder.

$$\emptyset_m^i = (1 \pm n)\emptyset_i \quad (6)$$

As mentioned, the discretization allows sparsity in the implementation. However, the field homogeneity is compromised with the sparsity. For MRI when traditional Fourier MR imaging method with linear gradient fields are used, an inner-field dipolar Halbach cylinder (n = 1) can be used. For such a Halbach cylinder, $\emptyset_m^i = 2\emptyset_i$, based on (6). When homogeneity is required, the imaging volume is relatively small, e.g. a volume of 3 x 3 x 5 $mm^2$ in [36] and 5 x 5 x 5 $mm^3$ for a homogeneity of less than 0.1 mT/0.311 T and 18 x 18 x 30 $mm^3$ with a homogeneity of 700 ppm in [43]. Shimming methods such as adding small magnetic shimming blocks inside the cylinder were proposed [44, 45].

### II.c Permanent Magnet Arrays for MR Imaging with Less Homogeneous $B_0$

In recent years, progress has been made to use non-linear gradient fields for MRI imaging [15, 21-24] . This part of research work was motivated by the need to overcome the physiological limitation of a traditional linear gradient field, e.g. peripheral nerve stimulation caused by linear gradient fields [15] [10]. The main idea is to encode an MRI signal using both a patterned magnetic field $\boldsymbol{B_0}(\boldsymbol{r})$ and coil sensitivity $C_q(\boldsymbol{r})$ (the $q^{th}$ coil of a coil array) where $\boldsymbol{r}$ is the position vector in the imaging domain. The signal of the $q^{th}$ coil in time domain can be expressed as follows [16],

$$S_q(t) = \int_V C_q(\boldsymbol{r})\, e^{-i2\pi\gamma \boldsymbol{B_0}(r)t} m(\boldsymbol{r}) d\boldsymbol{r} \quad (7)$$

$m(\boldsymbol{r})$ is the image and $\gamma$ is the gyromagnetic constant. In (7), $C_q(\boldsymbol{r})$ encodes the amplitude of the signal and $\boldsymbol{B_0}(\boldsymbol{r})$ is the SEM encoding the phase. Theoretically, $\boldsymbol{B_0}(\boldsymbol{r})$ can be of an arbitrary pattern. In the literature, a SEM is generated by coils (electromagnets) with a desired pattern, e.g. a pair of orthogonal multipolar fields. By using these methods, the reliance on traditional Fourier image encoding with linear gradient fields is reduced and the constrain on the homogeneity of $\boldsymbol{B_0}$ is relaxed. As a result, more permanent magnets/magnet arrays of different types with reduced magnetic materials and reduced weights, can therefore be used for MR imaging for an increased imaging volume, besides those provide homogeneous fields that are bulky and heavy, which may lead to a construction of a body scanner with portability and a low cost. A permanent magnet/magnet array supplies a relatively inhomogeneous magnetic field. The field is used as a spatial SEM serving as a combination of a $\boldsymbol{B_0}$ field and the gradient fields. The field pattern affects the method of image reconstruction as well as the quality of the reconstructed image (spatial resolution and SNR), whereas the field inhomogeneity decides the bandwidth of the RF components in an MRI system. For the NMR/MRI system that works with a less homogeneous $\boldsymbol{B_0}$ as SEM, broadband excitation, broadband RF coils, and spin-echo refocusing are needed. Next, the permanent magnets/magnet arrays that can be used or have been used to supply SEM for imaging are reviewed.

The magnets/magnet arrays that supply SEM for imaging can be categorized based on the relative location of the imaging volume/FoV and the magnet/magnet array. *ex-situ* magnet arrays are those that have the FoV outside the array, such as the magnet array for an inside-out NMR sensor and a U-shaped magnet/magnet array, whereas the *in-situ* cases are those that have the FoV inside the array, such as a Halbach array and an Aubert ring pair. To supply a homogeneous magnetic field, due to the stringent requirement on the homogeneity, the magnets/magnet arrays, as introduced in Section II.b, are *in-situ*. When the requirement on homogeneity is relaxed, both types of magnets/magnet arrays can be used. Next, these two types of magnets/magnet arrays will be reviewed in c1 and c2, respectively.

### c1. Permanent Magnet Array for Imaging Outside the Array, *ex-situ* Magnet Array

***ex-situ* magnet arrays** supply magnetic field for imaging outside the array. They are widely used in single-sided NMR devices. They supply less homogeneous $\boldsymbol{B_0}$ as 1D SEM for spatial encoding for 1D depth profiling. For these devices, the sample size is much bigger than the size of the magnet/magnet array which normally has an open geometry. An early example is the **inside-out NMR sensor** that is suitable for the cylindrical boreholes of oil wells in the oil well-logging industry [46]. Fig. 12 shows the inside-out well-logging NMR sensor designed by J. A. Jackson. As shown, a sensitive volume of homogeneity in a shape of an annulus is generated by two cylindrical magnets facing each other with the same poles at the center. The gradient field in the volume is 0.1 T/m. Electromagnet was used in Jackson's design while with the progress in the rare-earth permanent magnets [23], more effective inside-out sensors were developed using permanent magnet [47]. This type of design was extended to the design of a cardiovascular endoscope developed in Israel [48, 49] for bio-medical applications.

*ex-situ* magnet arrays are popular to supply **unilateral magnetic fields** for **unilateral NMR**. The magnetic field falls off with depth providing the gradient field for imaging. It can be as simple as **a bar magnet** as shown in Fig. 13 (a) where the RF coil is place on one of the faces. In this case, $\boldsymbol{B_0}$ is perpendicular to the pole face and the sensitivity volume scales with the size of the pole face. Besides a single bar magnet, there are other variants making use of this type of perpendicular unilateral magnetic field for imaging. One example is **the barrel magnet** where two concentric tube magnets with different diameters and opposite polarization, or a magnet with a drilled hole is used [50]. For a barrel magnet, a sweet spot where the gradient field is zero can be generated.

Another classic structure for unilateral imaging is the U-shaped magnet (or called horseshoe magnet) as shown in Fig. 13 (b). As shown in Fig. 13 (b), $\boldsymbol{B_0}$ field goes approximately horizontally (with a quadratic profile along the z- and the x-direction) from one pole to the other, and the RF coil is located between the two magnets supplying a vertical $\boldsymbol{B_1}$ field. For the magnet array of this type, $\boldsymbol{B_0}$ field is in parallel with the two pole faces and the sensitive volume scales with the size of the gap. An **NMR MOUSE** (**mo**bile **u**niversal **s**urface **e**xplorer) was proposed for non-destructive investigation of arbitrarily large objects making use of the unilateral field and gradient of a U-shaped magnet [51]. It was the first truly portable NMR system. Lately, it was applied to profile human skin in one dimension [52]. For a basic U-shaped magnet, the magnetic field has a curvature on the surface between the poles. The curvature limits the depth resolution to a few points across an accessible depth. Approaches were proposed to eliminate the curvature by changing the shape of the magnets [53], rearranging the magnets [54], or adding shimming blocks between the two main poles [55].

Recently, a 'cap-like' magnet array was proposed to use unilateral magnetic field for head imaging [56]. An equatorial portion of Halbach sphere was optimized and implemented using 1"x1"x1.375" magnet blocks with the magnetic field pointing on the equatorial plane and a gradient field along the center axis.

**c2. Permanent Magnet Array for Imaging Inside the Array, *in-situ* Magnet Array**

***in-situ* magnet arrays** supply magnetic field for imaging inside the array. The well-known *in-situ* magnet arrays are Halbach cylinder [32, 33] and Aubert ring pair [57]. To supply 2D SEM for MRI, dipolar Halbach cylinder and dipolar Aubert ring pair(s) can be used.

**c2-1 Halbach Array**

As introduced in section II.b2, a Halbach cylinder supplies dipolar magnetic fields pointing on the transversal direction of the cylinder. It was proposed to be used for MRI imaging based on traditional Fourier imaging approach with linear gradient fields [36, 43]. In the reported work, because a high homogeneity is needed, the imaging volume of the proposed structure is in the scale of less than 2 cm in length. When the magnet field is used for both the main field and the gradient fields, the constrain on the field homogeneity can be relaxed, and it allows an increased imaging volume, such as one for head imaging.

In [20], a sparse dipolar Halbach cylinder (an "NMR Mandhala") was used to supply $\boldsymbol{B_0}$ for head imaging. It consists of 20 N42 grade NdFeB magnet bars (1"x1"x14") and two shimming rings each of which consists of 20 1" NdFeB cubes at the end of the cylinder. The magnet array weights 45 kg and has an average field of 77.3 mT in a center plane with a diameter of 16 cm. Fig. 14 (a) and (b) shows the 3D view and side view of this Halbach cylinder, respectively. Fig. 14 (c) and (d) are the measured magnetic field on the yz- (transversal) and the xy-plane (longitudinal). The field pattern on the yz-plane is the SEM for imaging. As shown in Fig. 14 (c), the field pattern is quadrupolar (with a dipolar direction).

In this head imaging system in [20] with the quadrupolar SEM, the mapping between the object space and the encoding space is nonbijective, which leads to aliasing in the image through the origin. To disambiguate the nonbijective mapping, in [20] SEM is rotated, and stationary encircling receive coils were used to encode the amplitude of the signal. The signal equation of this system is expressed as below,

$$S_{q,\theta}(t) = \sum_{\boldsymbol{r}} C_q(\boldsymbol{r})\, e^{-i2\pi\gamma \boldsymbol{B_0}(r,\theta)t} m(\boldsymbol{r}) \quad (8)$$

where $S_{q,\theta}(t)$ is the signal received from the $q^{th}$ RF coil in time domain when the SEM is rotated at an angle of θ, $m(\boldsymbol{r})$ is the image, $\boldsymbol{r}$ is the position vector in the imaging domain, $C_q(\boldsymbol{r})$ is the sensitivity of the $q^{th}$ coil, $\boldsymbol{B_0}(\boldsymbol{r},\theta)$ is the distribution of the static magnetic field (the SEM) at angle θ for spatial encoding, and $\gamma$ is the gyromagnetic constant. For image reconstruction, the following system can be constructed,

$$\boldsymbol{S} = \bar{\bar{E}}\boldsymbol{m} \tag{9}$$

where $\boldsymbol{S}$ is a vector consists of signals from the RF coils that are measured at different rotation angles, $\bar{\bar{E}}$ is the encoding matrix, and $\boldsymbol{m}$ is the image that is arranged into a vector. To solve this system, the iterative method, Kaczmarz method [58] was used. Recently, the system was solved by using truncated single value decomposition (TSVD) which has a reduced computational complexity [59]. In this approach, the information on the image is preserved whereas the noise is truncated. The TSVD method shows significant improved image quality when the system is noisy compare to the iterative method. Besides, a short Halbach cylinder was reported with 2D field of view of 12 cm in diameter [60]. Two shimming rings were used to mitigate the end field effect of the cylinder.

When the magnetic field from a magnet array serves as the SEM which is a combination of both $\boldsymbol{B_0}$ and gradient fields, the advantage is that it removes the need for gradient coils. However, on the other hand, the field strength is lower compared to those when a SEM is applied to a superconducting magnet-based system with a high magnetic field (e.g. > 1.5T). Moreover, the homogeneity of the field in the FoV need to be controlled in the range that the RF excitation and RF coils can handle. With the concerns above, the sparse Halbach cylinder in [20] is further optimized in terms of the field strength and field homogeneity for imaging using Generic Algorithm (GA) [61].

**c2-2 Aubert Array**

An Aubert ring pair consists of two annular magnets of the same dimension with the central axes aligned and located a distance apart, forming a cylindrical space. Fig. 15 shows the side view, front view, and end view of the ring pair. As shown, in the ring pair, one magnet ring has the magnetization radially pointing outward and the other radially pointing inward. It supplies dipolar magnetic field along the axial direction of the cylinder. It was invented by G. Aubert and reported in his patent in 1994 [57]. An Aubert ring pair may be applied to MRI as $\boldsymbol{B_0}$ with linear gradient fields when the field is homogeneous, and as 2D SEM when the field become less homogeneous in an increased FoV.

In [62], an Aubert ring was implemented by discretizing the rings and using magnet cubes to build it. Fig. 16 shows an illustration of the implemented Aubert ring in [62]. As shown, there are two rings at each side and each ring consists of 12 magnet cubes (12 mm x 12 mm x 12 mm). The bore size is 52 mm in diameter and the total size is 66 mm x 90 mm x 90 mm, with a weight of 1 kg. It provides a field strength of 120 mT at the center and homogeneity of 15 ppm over a sphere of 5 mm in diameter for traditional imaging using linear gradient fields. The field can be further shimmed by introducing more rings [63].

When the FoV increases, the field homogeneity is lowered. Although non-Fourier imaging methods without linear gradient fields can be applied to a system when such a magnet array is used to supply the magnetic field, when the FoV is increased to accommodate a human head, the field strength is further weaken while the field homogeneity is further lowered, which become too low to be tolerated by the RF coils and RF control circuit, and severely lower the SNR. In [64, 65] , a ring-pair permanent magnet array based on an Aubert ring pair, an Aubert ring-pair aggregate, was proposed to supply SEM for head imaging with an optimized field strength and field homogeneity. Fig. 17 (a) and (b) show the 3D and side view of the proposed ring aggregate. As shown in Fig. 17, it consists of *n* Aubert ring pairs with the same outer radius and optimized inner radius. The distance between the two rings of the inner most pair is *d*. GA was used to optimize the inner radius and the distance *d* when the outer radius is set to be 250 mm. 10 pairs of rings are used. The FoV is a cylinder with a diameter of 200 mm and a length of 50 mm. Fig. 18 (a) shows the flow of optimization using GA. The objectives are high field strength (> 160 mT) and high field homogeneity (< 20000 ppm). To accelerate the optimization, a fast forward calculation was proposed [65] where the calculation of the magnetic field of the magnet rings is speeded up by applying axial symmetry to a current model. Fig. 18 (b) shows the optimized inner diameters of the ring-pair aggregate.

Fig. 19 and Fig. 20 show the magnetic field of the optimized magnet array and that of the original Aubert ring pair with the same mass and a similar dimension, respectively. As shown in these two figures, in the same FoV, the optimized magnet array has a field homogeneity of 24,786 ppm and an average field strength of 169.7 mT whereas the original Aubert ring pair has a field homogeneity of 122,150 ppm and an average field strength of 178.5 mT. As can be seen, the optimization reduces the inhomogeneity by over 79.7% with a sacrifice of the field strength of less than 5% (from 178.5 mT to 169.7 mT). The optimization offers a significant improvement in the field homogeneity while still maintains a similar field strength. The optimized array satisfies one of the pre-set optimization objectives which is an average field strength of less than 160 mT, and it provides a field homogeneity of 24,786 ppm, which is very close to the other objective, a field homogeneity of less than 20000 ppm.

To facilitate an implementation, the proposed design is segmented into 12 identical fan-shaped magnets with a uniform polarization. The 3D view of the magnet assembly is shown in Fig. 21 (a). The segmented magnet assembly was simulated in COMSOL Multiphysics. The simulation results are shown in Fig. 21 (b) and (c). The segmented magnet array has a field homogeneity of 32,511 ppm and an average field strength of 167.6 mT. Comparing to the results of the optimized array with a continuously varying polarization in Fig. 19, owing to the discretization, the field homogeneity is reduced by 31.2% and the field strength is lowered by 1.24%.

### II.d Other Aspects of Permanent Magnet Arrays

Due to the decaying of magnetic field away from a permanent magnet, the magnet shows a natural gradient in its proximity. When an array is formed, a field pattern can be expected. Field patterns can be applied as SEM for MRI imaging where 1D and 2D examples were reported in the literature. Moving forward, *ex-situ* or *in-situ* permanent magnet arrays can be designed as 3D SEM for imaging.

For the assembly of a magnet array, due to the inevitable imperfection of magnet materials and fabrication, there is non-negligible difference among the magnets in terms of remanence strength, and even the direction of polarization. The effect of imperfect materials and fabrication should be identified and minimized. One way to minimize the effect of material imperfection is labelling, measuring, and sorting the magnets before assembling [62][54] thus those with serious defects could be discarded, and the arrangement of magnets could be optimized. Moreover, there are discrepancies between a defined location of a magnet and its real location. The accuracy of the position of the magnet can be increased if a housing is properly designed and fabricated with grooves of right dimensions for the magnets. To fabricate such as housing structure, 3D printing provides a fast and affordable option.

Another downside of permanent magnets is the slow degradation of remanence over time. However, with the progress on magnet manufacture, the stability on magnetization has been significantly improved based on observations in the recent experiments on high resolution NMR spectroscopy. Besides, sensitivity of magnetic materials to temperature variation is another issue to permanent magnets, which means that the magnetic field varies when temperature changes. The temperature sensitivity of a magnetic material is quantified using its temperature coefficient, which is $(\Delta B/B)/\Delta T$. The commonly used magnetic material, NdFeB, has a large temperature coefficient, -1.1 $\times$ $10^{-3}$/°C. In order to maintain a stable magnetic field, the most straightforward solution is to have a relative thermostatic environment. When temperature fluctuates within a certain range, approaches have been proposed in the literature to compensate the change of the magnetic field so as to obtain a thermal stability. A common approach for temperature compensation is to use another compensative alloy that have a designed temperature coefficient and size to compensate the change and to maintain a relatively stable resultant magnetic field in a defined temperature range [66-68]. Alternatively, the change of a field caused by thermal fluctuation can be compensated using electromagnets [69]. Another possible solution to this problem is to add an NMR probe to the system to track the drift of the field strength and compensate it during post processing [20].

## III. Discussions and Conclusion

Body MRI has been available for about four decades. It is dominated by imaging method based on Fourier transfer which requires a homogeneous static magnetic field ($\boldsymbol{B_0}$) and linear gradient fields in the FoV. To supply the static magnetic field with homogeneity, a big magnet (superconducting magnets, electromagnets, or permanent magnets) is needed which makes an MRI system bulky and expensive. As a result, MRI is not widely available when the space to sit it is small, or in the environment when electric power and other resources are limited. It is not available to underdeveloped countries and areas, or to routine scans either, due to a high cost. An MRI body scanner with portability (with a size and weight that can be handled by a trolley or a man), reasonable power consumption, and a low cost can definitely enable the “point-of-care” and time-sensitive imaging and diagnosis. Meanwhile, with a reduced cost, it can make this imaging modality available to more people who need it and those who need it regularly.

Permanent magnet arrays with sparsity, a reduced size, and a reduced weight can be used to supply spatially non-linear encoding static field for imaging in a relatively large FoV. Because of the simplicity and low cost of permanent magnets, this offers a possibility to construct a portable and low-cost MRI body scanner. Compared to superconducting magnets and electromagnets, permanent magnets do not require a cooling system and they do not have power consumption. NdFeB rare-earth magnets have high energy products and supply relatively high magnetic fields at room temperature compared to other permanent magnets. Different configurations of magnetization were proposed to supply specific field patterns. A well-known one is the Halbach array where flux is guided on one side of the array and it can be zero on the other side. This offers a chance for an MRI system to be sit in a small space, such as an ambulance, without a concern on de-metaling in the region outside the imaging volume. *in-situ* magnet arrays, e.g. a Halbach cylinder and an Aubert ring pair/ring-pair aggregate, supply dipolar fields inside the array for MR imaging in the cylinder-transversal direction and the cylinder-axial direction, respectively. The *ex-situ* counterparts supply magnetic fields outside the arrays, either perpendicular or parallel to the pole faces. The field inhomogeneity increases when the FoV increases for body imaging, and when the arrays become sparse to reduce the weight and to facilitate fabrication. The field patterns that can be applied as a SEM act as a combination of the $\boldsymbol{B_0}$-field and the gradient fields. With this approach, the need for gradient coils can be removed. Therefore, the system can be further simplified, and the power consumption can be further reduced. The field patterns can be non-bijective. In this case, making use of the sensitivity of receiving coils for further encoding helps to eliminate aliasing[15, 20].

In the literature, *ex-situ* and *in-situ* permanent magnet arrays are designed as 1D and 2D SEM. Taking a similar approach, they can be designed as 3D SEM for body imaging in the future. Optimization tools such as GA, particle swarm optimization (PSO) etc. can be applied. It should be noted here that the forward calculation in an iteration for optimization must be fast for an efficient optimization. The calculation can normally be accelerated based on the specific geometry of the structure being optimized. For example, an axial symmetry was used to accelerate the forward calculation when an Aubert ring-pair aggregate was optimized [64, 65]. To tailor the field pattern, small magnets, small iron bars/cubes, and electromagnets can be helpful.

When permanent magnets/magnet arrays and SEM are applied together to build a portable and low-cost MRI scanner, the average field strength is low (< 1 T), which results in a low SNR and a low resolution. By taking this approach, the SNR and resolution are not uniform. They are affected locally by the field pattern. Owing to the limited resolution, such a portable and low-cost system, for now, may not be as useful as a diagnostic tool as a conventional MRI scanner. However, it may offer in-time and at-site detection and diagnosis of acute medical situations, for example, time-sensitive posttraumatic brain haemorrhage, for patient management. For compensating a low SNR, advanced physical concepts, such as metasurface, could be a good remedy to apply to redistribute the $\boldsymbol{B_1}$-field of receive coils, in ways to increase the coil sensitivity and to increase the penetration depth [70, 71]. Besides, to increase the SNR, staying away from pulsed excitations, approaches of using continuous waves to simultaneously transmit and receive RF signals can be helpful [79, 80].

Using SEM for imaging, it is noted that the inhomogeneity of the field is related to the working bandwidth of the radiofrequency (RF) excitation and RF coils. The required bandwidth increases as the inhomogeneity increases, which leads to a decrease in Q-factor. A decreased Q-factor will further lower the SNR of the system using permanent magnets/magnet arrays. With this perspective, a wideband excitation, wideband RF coils, and a wideband matching without scarifying the transmission efficiency or reception sensitivity are needed.

The power consumption is another key factor to a portable system where a relatively low power consumption may enable an ‘off-the-grid’ operation, so as to make truly ‘point-of-care’ imaging possible. As analysed, using permanent magnets/magnet arrays eliminates the power consumption for the supply of $\boldsymbol{B_0}$. Furthermore, when the magnetic field supplied by magnets/magnet arrays has a patterned and is applied as SEM for signal encoding, the power consumption for gradient coils is eliminated since SEM is a combination of $\boldsymbol{B_0}$-field and gradient fields and no gradient coils are needed. In terms of reducing the RF peak-power, an RF system with simultaneous transmit and receive signals [72, 73] can be used.

For the computational complexity of the algorithms for image reconstruction using SEM, it is decided by the size of the encoding matrix, $\bar{\bar{E}}$, which is determined by the number of pixels/voxels, the number of receive coils, and a few more factors which depend on the specific encoding mechanism. For example, the number of rotating angles in the Halbach head imager [20, 21]. A direct inversion of $\bar{\bar{E}}$ is costly and it becomes computationally impractical in most imaging situations. Iteration methods such as the conjugate gradient method [74], Kaczmarz method [58], and TSVD [59] help to accelerate the calculation. A decomposition of $\bar{\bar{E}}$ into matrices, which are either sparse or there are fast algorithms exist for their inversion, was proposed to accelerate an image reconstruction [16].

A permanent magnet/magnet array shows great potential to construct a portable and low-cost MRI body scanner. In such a system, a permanent magnet/magnet array is used to supply SEM for signal encoding. The design of the permanent magnet/magnet array, the RF system (excitation, coils, shield, and matching), and the methods for image construction are closely associated together. They need to be optimized holistically to increase the SNR for good image quality. Away from the hardware, design of pulse sequence and post processing of image are important to the image quality which are not discussed in this content due to the different focus of the paper. In terms of post processing, deep learning can be applied with big data size and a fast searching speed to further improve the image quality possibly by learning from good-quality-bad-quality-image-pairs.

## Acknowledgement

The authors would like to thank their funding agency, Singapore- MIT Alliance for Research and Technology (Innovation Grant ING137068) and Singapore Ministry of Education (MOE Tier 1)

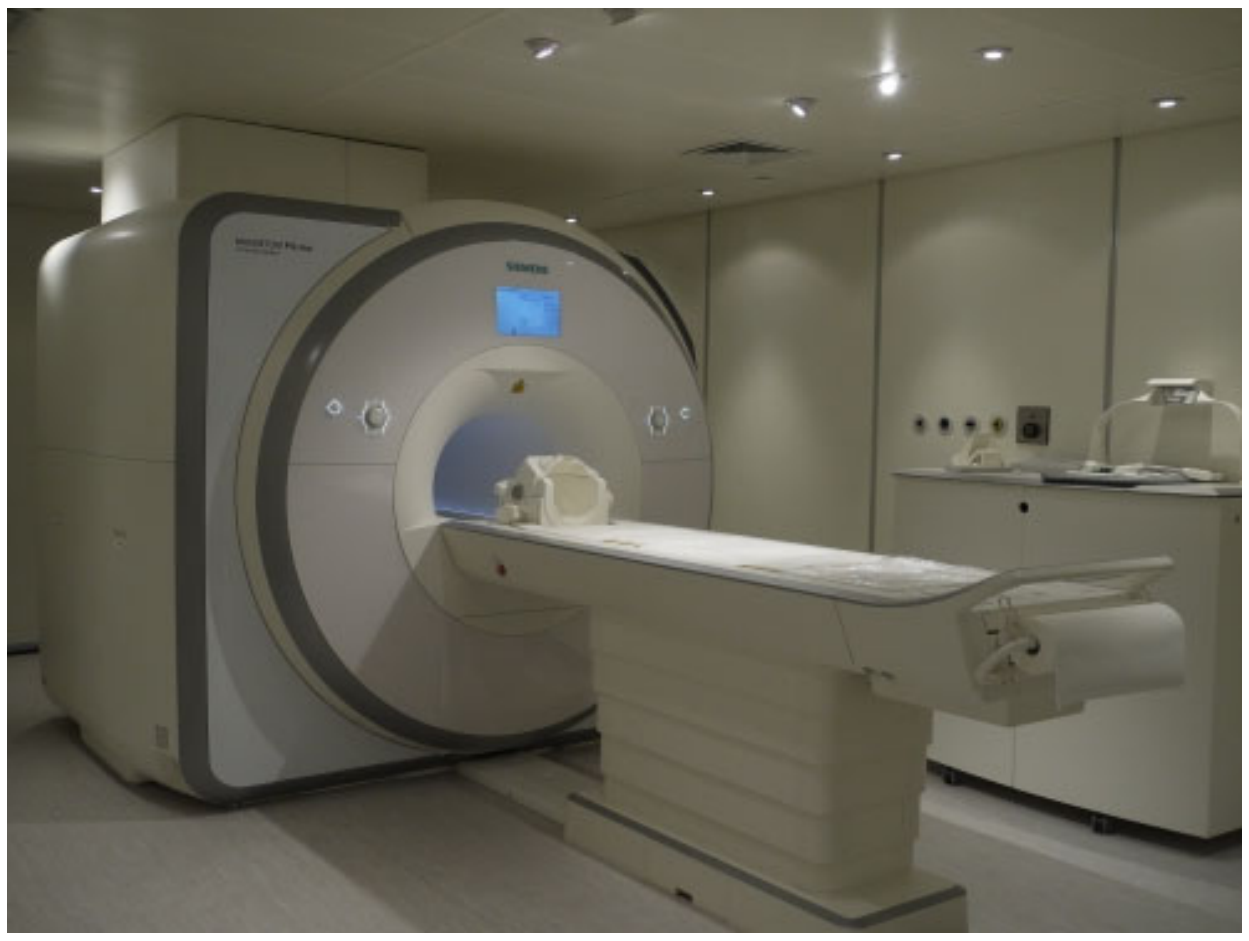

Fig. 1 Image of Siemens 3T MAGNETOM Prisma at A*STAR-NUS Clinical Imaging Research Centre (CIRC), superconducting magnet, 60 cm bore diameter, 213 cm long, system weight of 13 tons, minimum room size of 33 $m^2$, FOV of 50x50x50 cm, Gradients 80 mT/m @ 200 T/m/s [75]

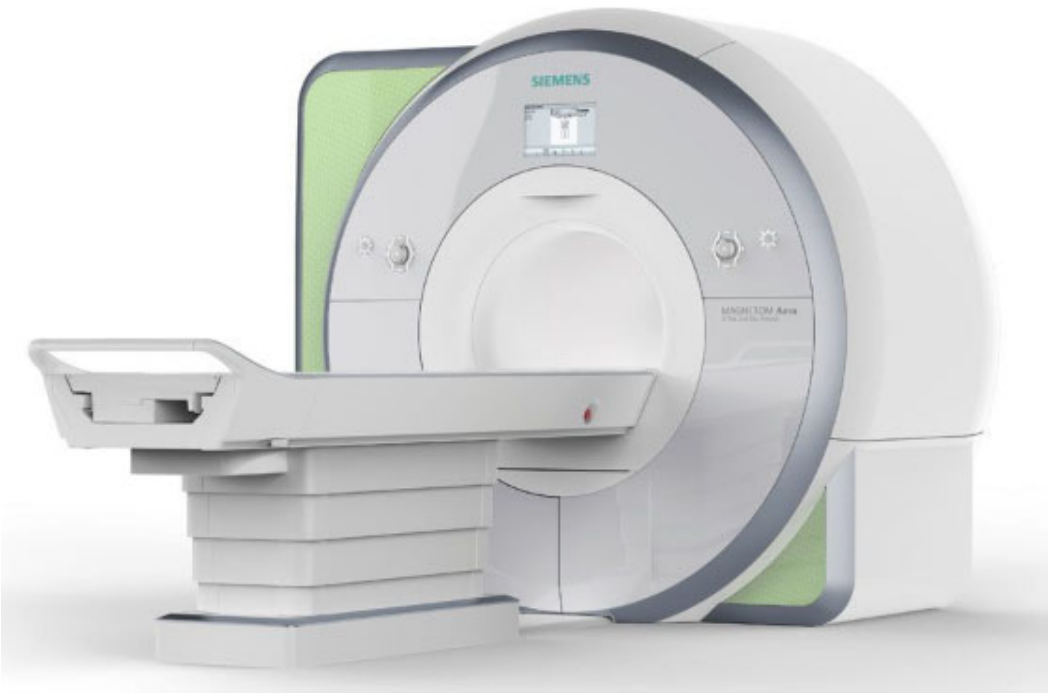

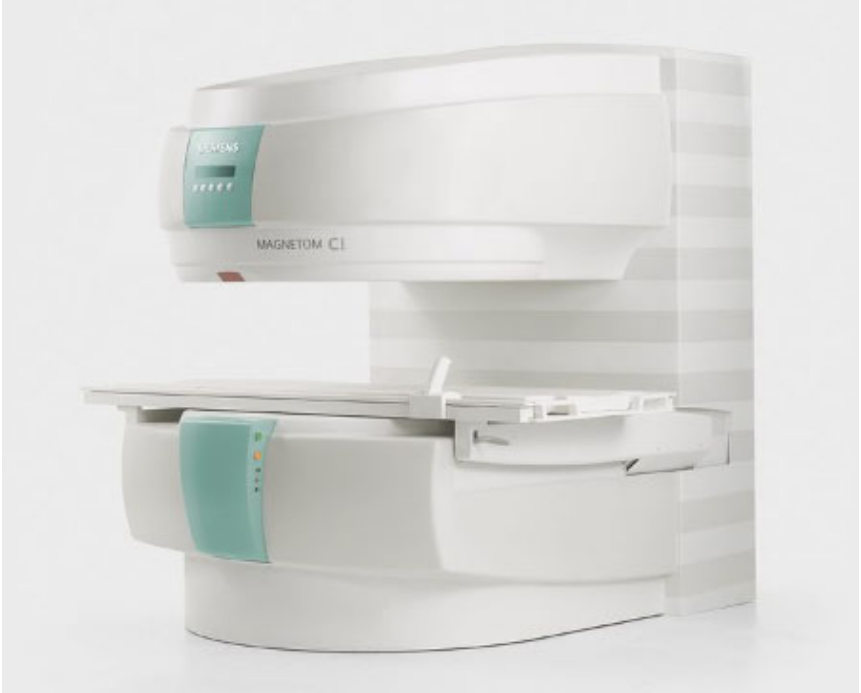

(a) (b)

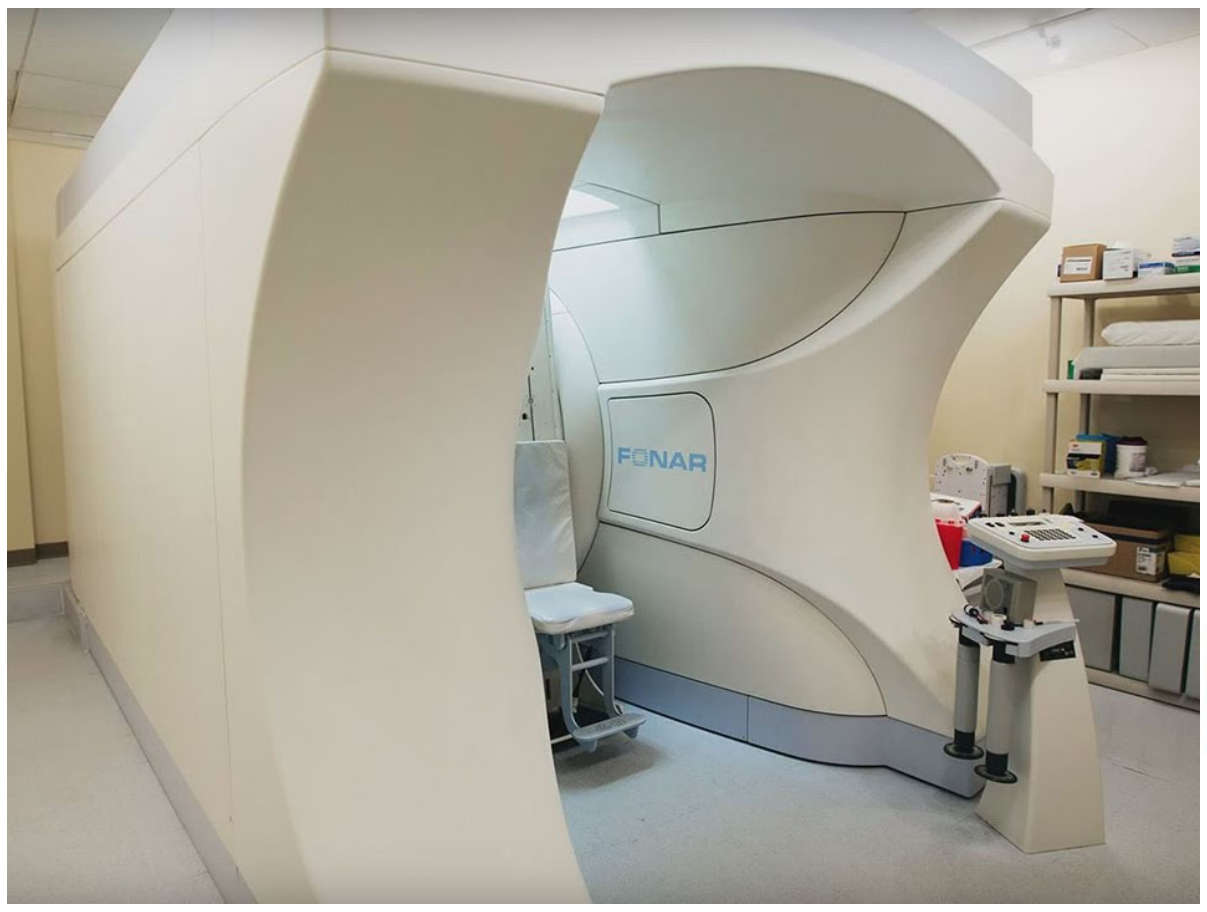

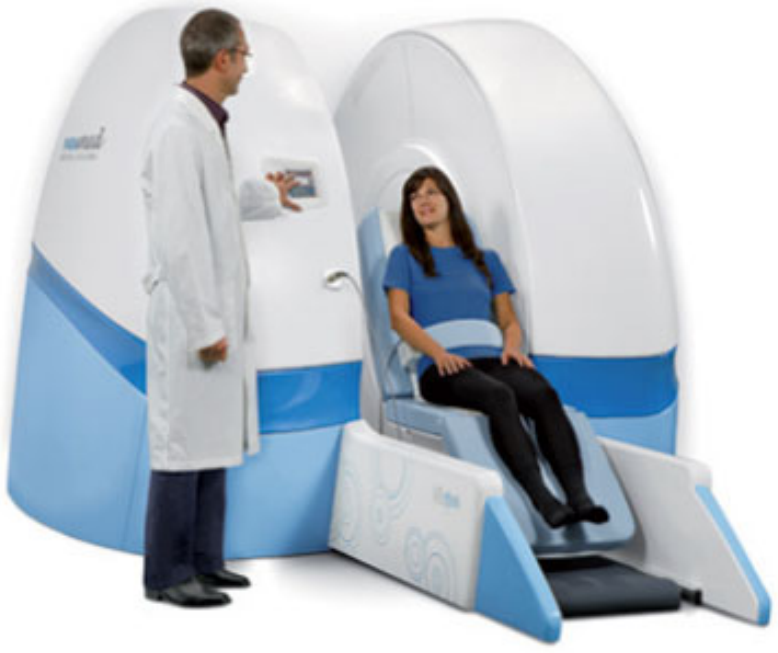

(c) (d)

Fig. 2 Examples of open MRI (a) Siemens 1.5 T MAGNETON Aera, short cylindrical superconducting magnet, 70 cm bore diameter, 137 cm long, system weight of 4.8 tons, minimum room size of 30 $m^2$, FOV of 50x50x50 cm, gradients 33 mT/m @ 125 T/m/s [4], (b) Siemens 0.35T MAGNETON C, C-shaped dipolar permanent magnet with a vertical magnetic field, bore gap size of 41 cm, $270^o$ accessibility, pole diameter of 137 cm, system weight of 17.6 tons, system dimension of 233 $\times$ 206 $\times$ 160 cm, minimum room size of 30 $m^2$, FOV of 0.5-40 cm, gradients 24 mT/m @ 55 T/m/s [5], (c) 0.6 T Upright$^{TM}$ MRI from Fonar, dipolar electromagnet with a horizontal magnetic field, bore gap size (pole-to-pole) of 46 cm, power requirement of 380-480V, FOV of 6 cm, gradients 12 mT/m, closed-loop water cooling, active and passive shimming, unreported system weight, system dimension, or minimum room size [6], (d) 0.5 T PARAmed open MRI, dipolar superconducting magnet using MgB2 with a horizontal magnetic field, cryogen free, low power consumption, bore gap size (pole-to-pole) of 46 cm [7].

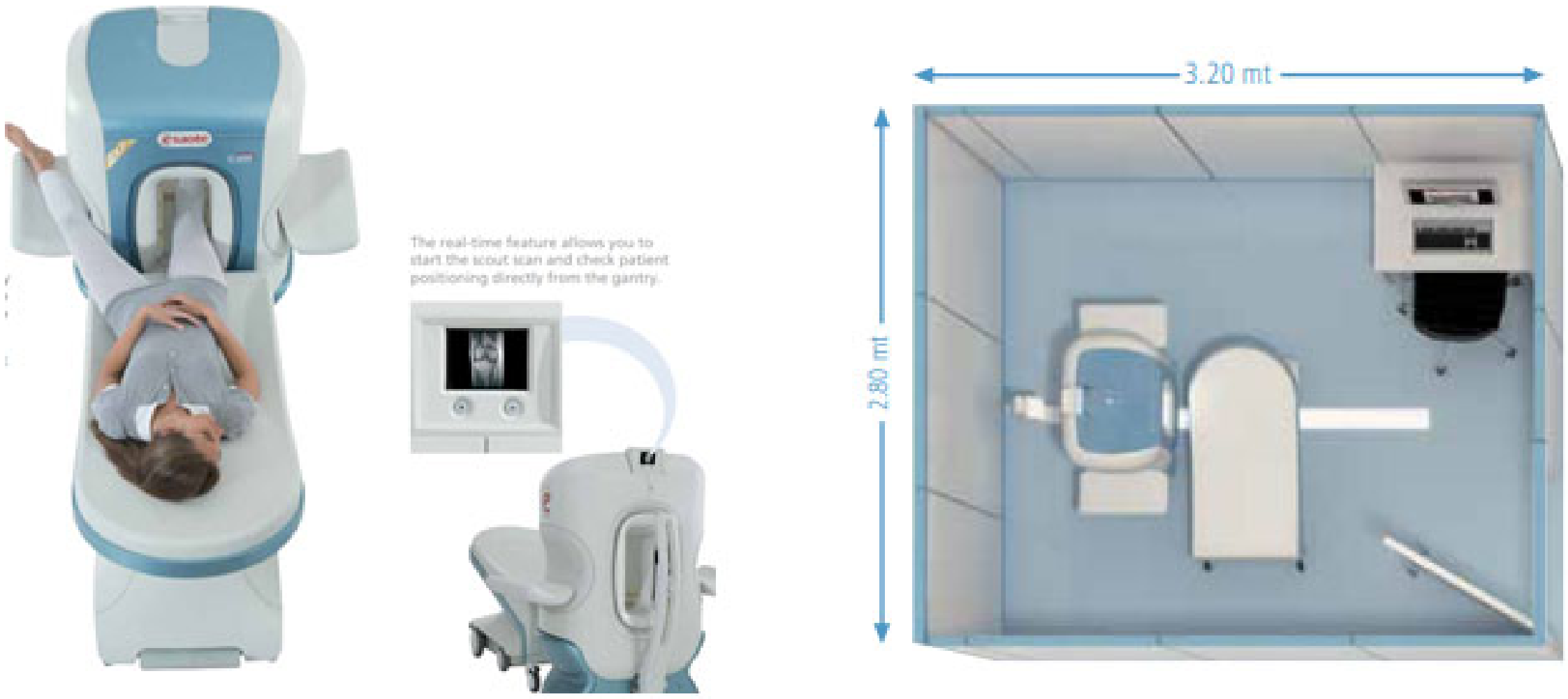

Figure. 3 O-scan from Esaote [9]

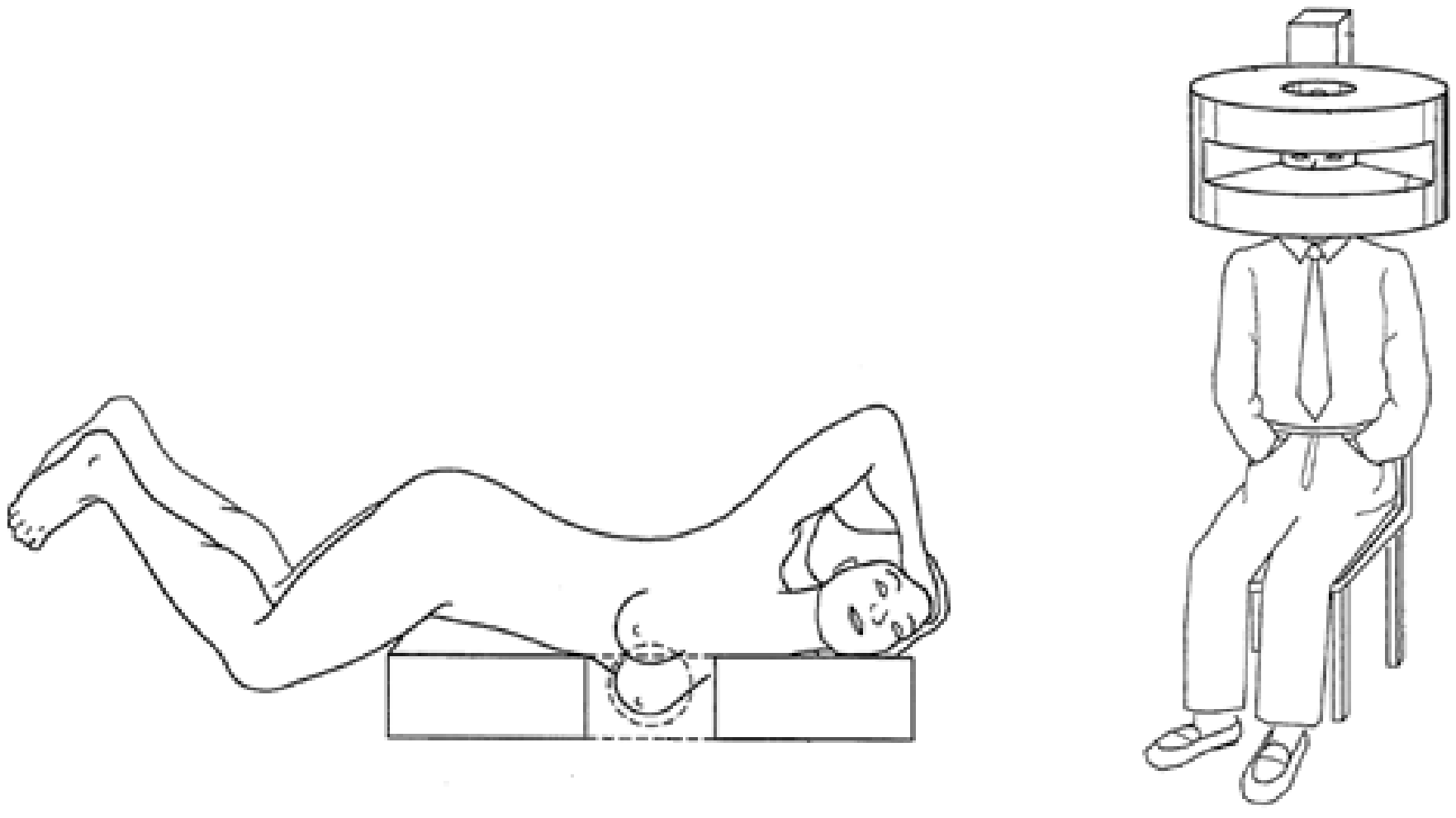

(a) (b)

Fig. 4 The organ specific superconducting magnets [10] (a) A bagel-shaped superconducting magnet for breast imaging (b) a helmet-shaped superconducting magnet for head imaging

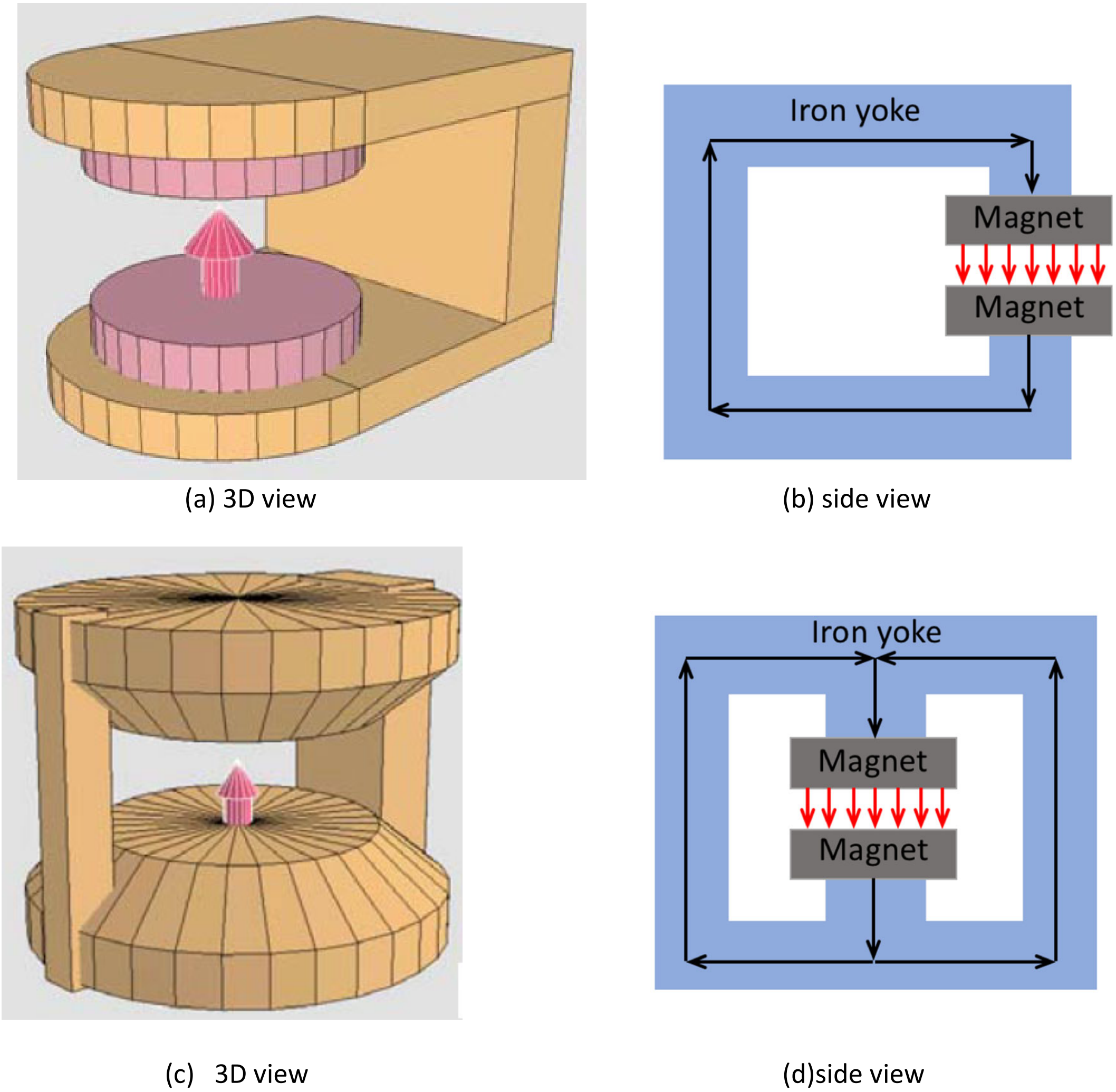

Fig. 5 Dipolar magnets (a) C-shaped, (b) H-shaped

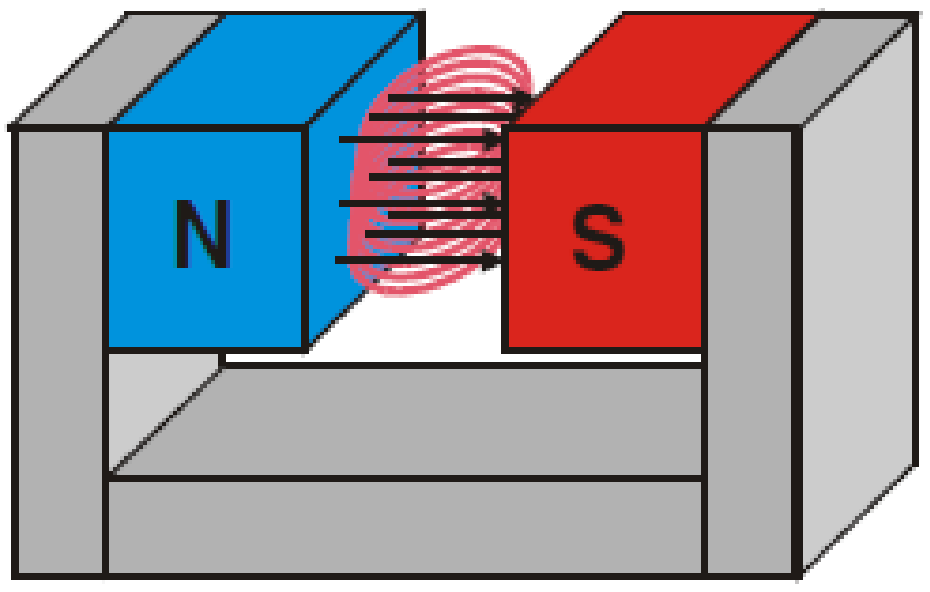

Fig. 6 A conventional C-shaped magnet

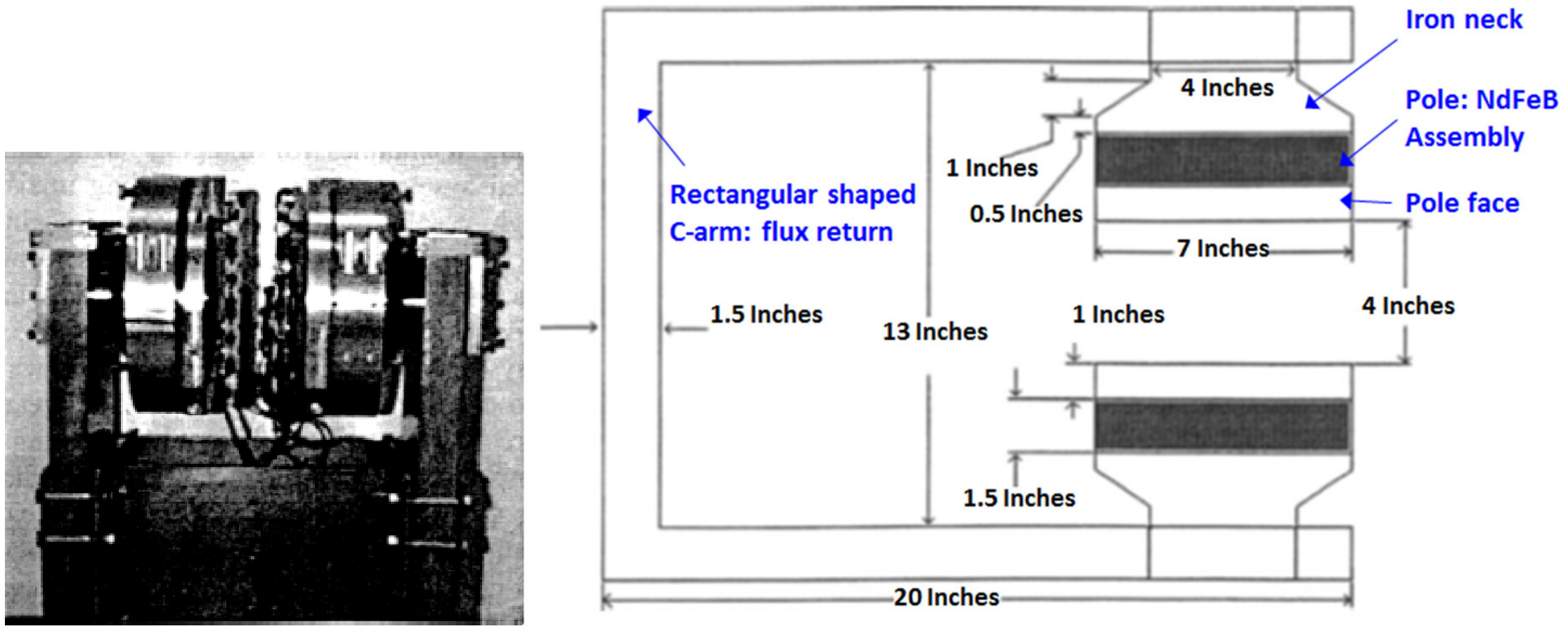

(a) (b)

Fig. 7 A C-shaped table-top permanent magnet array (0.21 T) for MRI imaging [12], (a) a photograph, (b) a side view of the system with dimensions

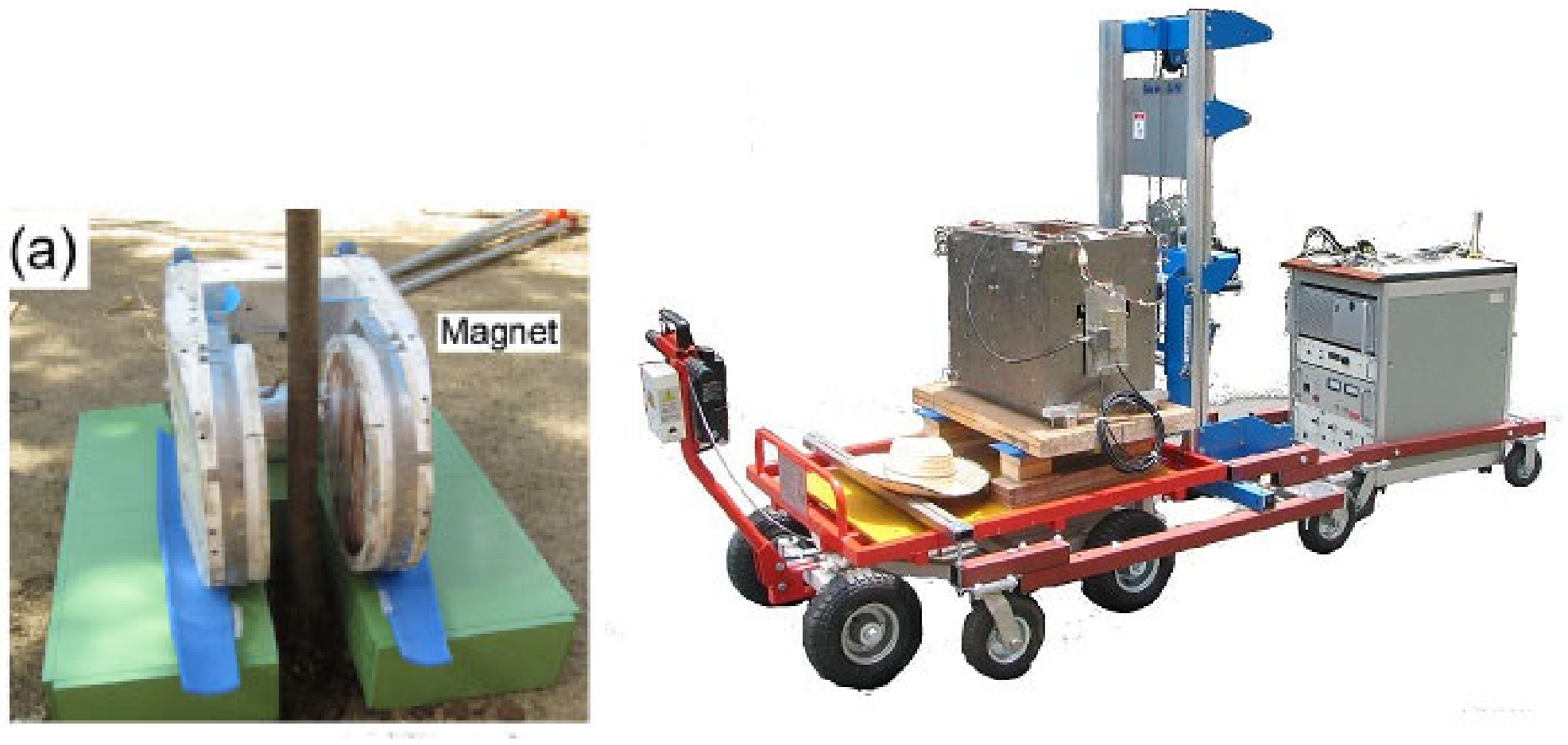

(a) (b)

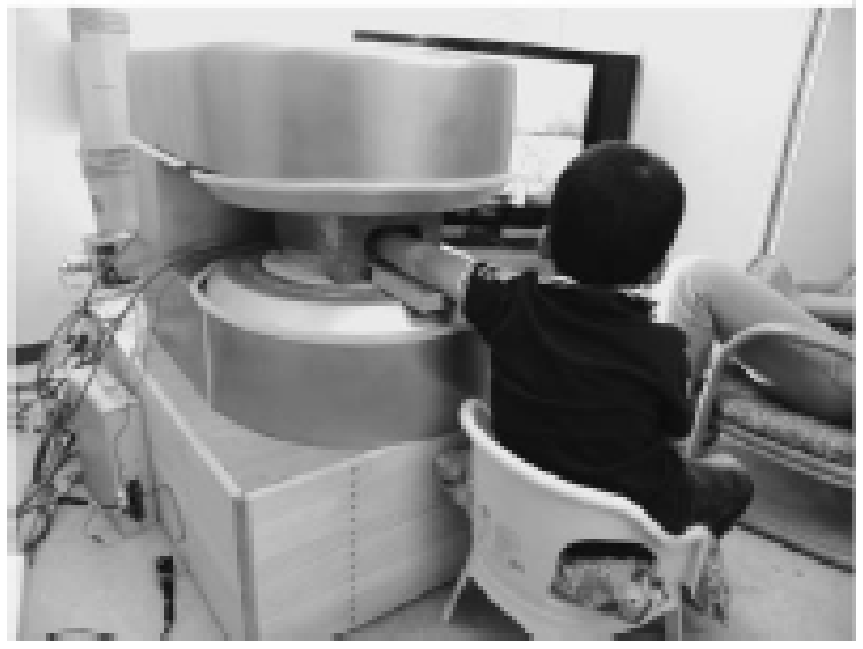

(c)

Fig. 8 C-shaped magnets for MRI, Terada Lab, Unversity of Tsukuba, (a) the magnet for scanning living trees (0.21 T) [28], (b) the MRI scanner for scanning living, (c) the magnet for scanning a hand (0.3 T) [29]

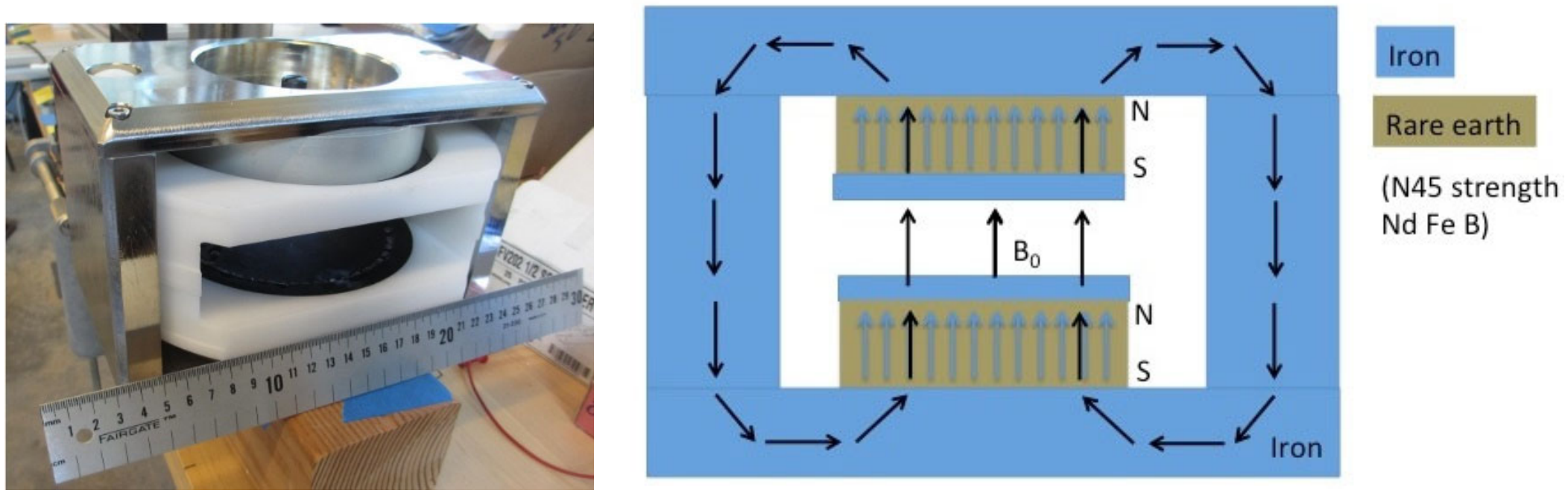

(a) (b)

Fig. 9 The magnet built by the Institute of Electrical Engineering of the Chinese Academy of Sciences in Beijing [30]

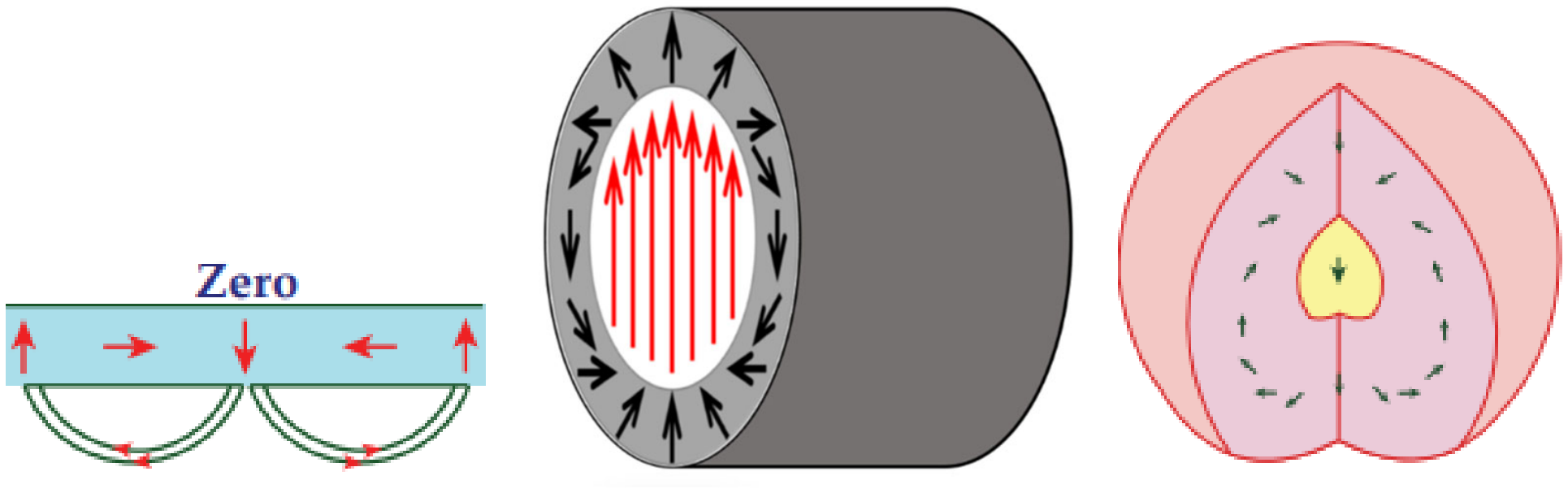

(a) (b) (c)

Fig. 10 Halbach permanent magnet array (a) 1D [34] (b) 2D (c) 3D [76]

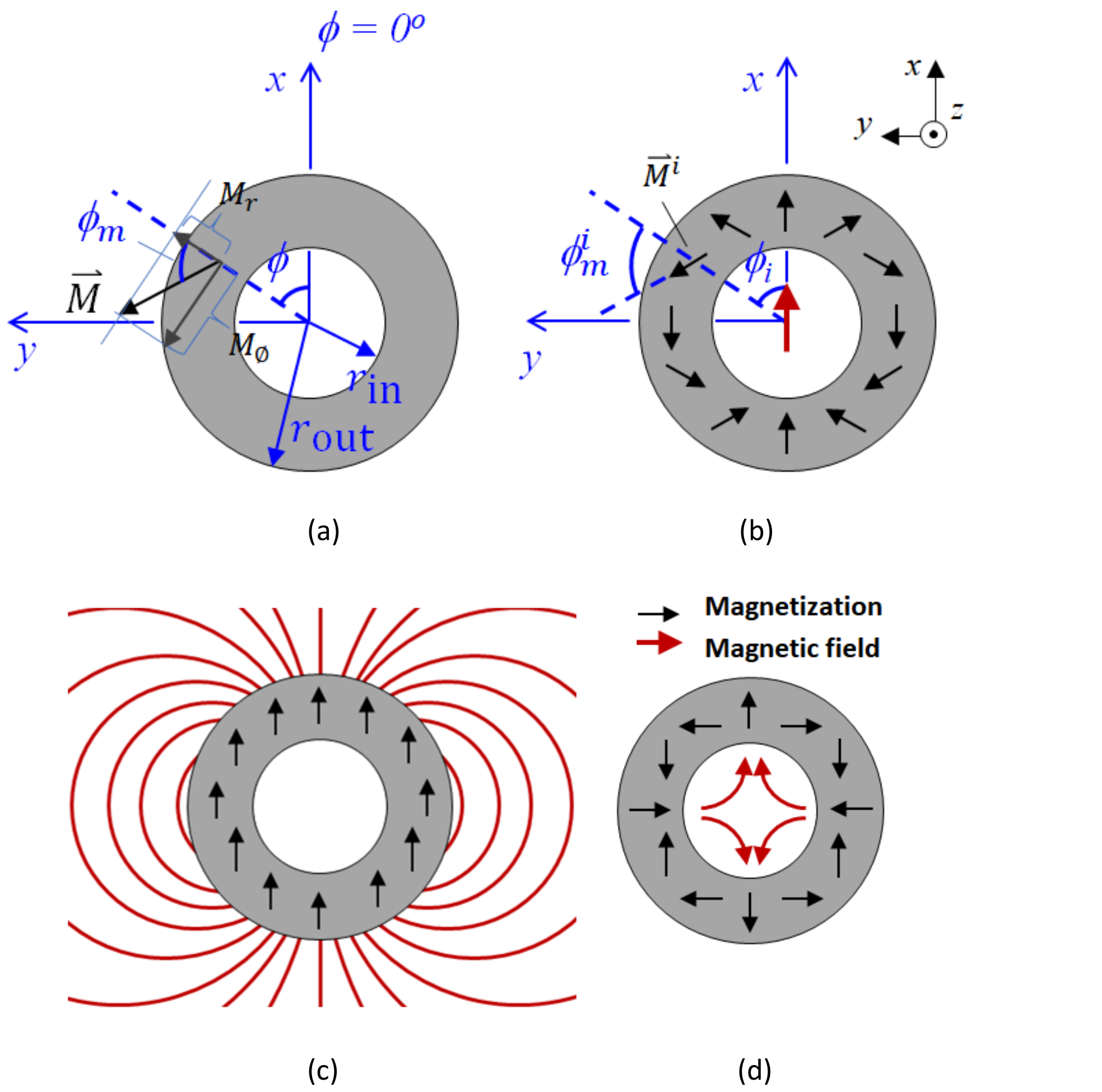

Fig. 11 Side views of a Halbach cylinder (a) relationship of $\vec{M}$ and angles, (b) inner-field, dipolar (n = 1), (c) outer-field, dipolar (n = 1) (d) inner-field, quadrupolar (n = 2)

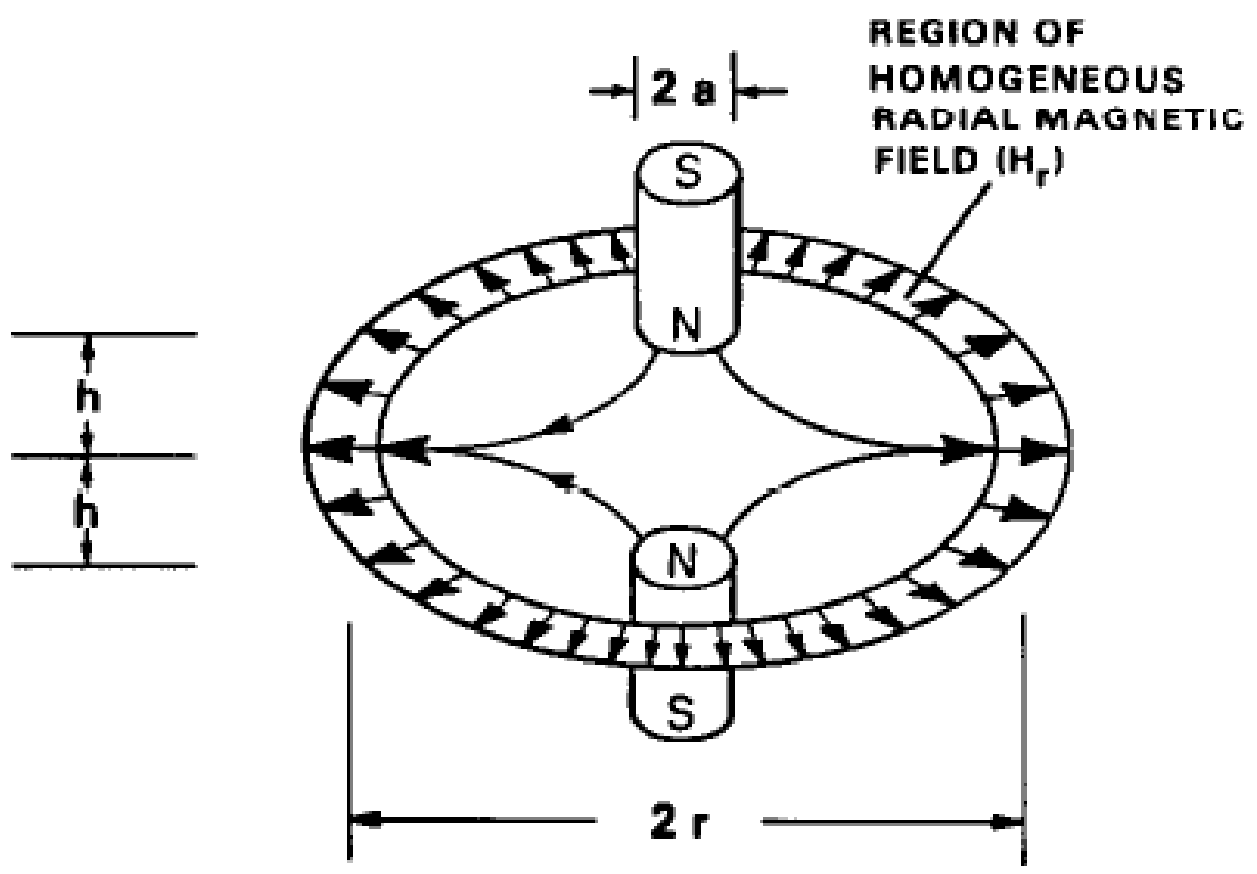

Fig. 12 The inside-out well-logging NMR sensor designed by Jackson [46]

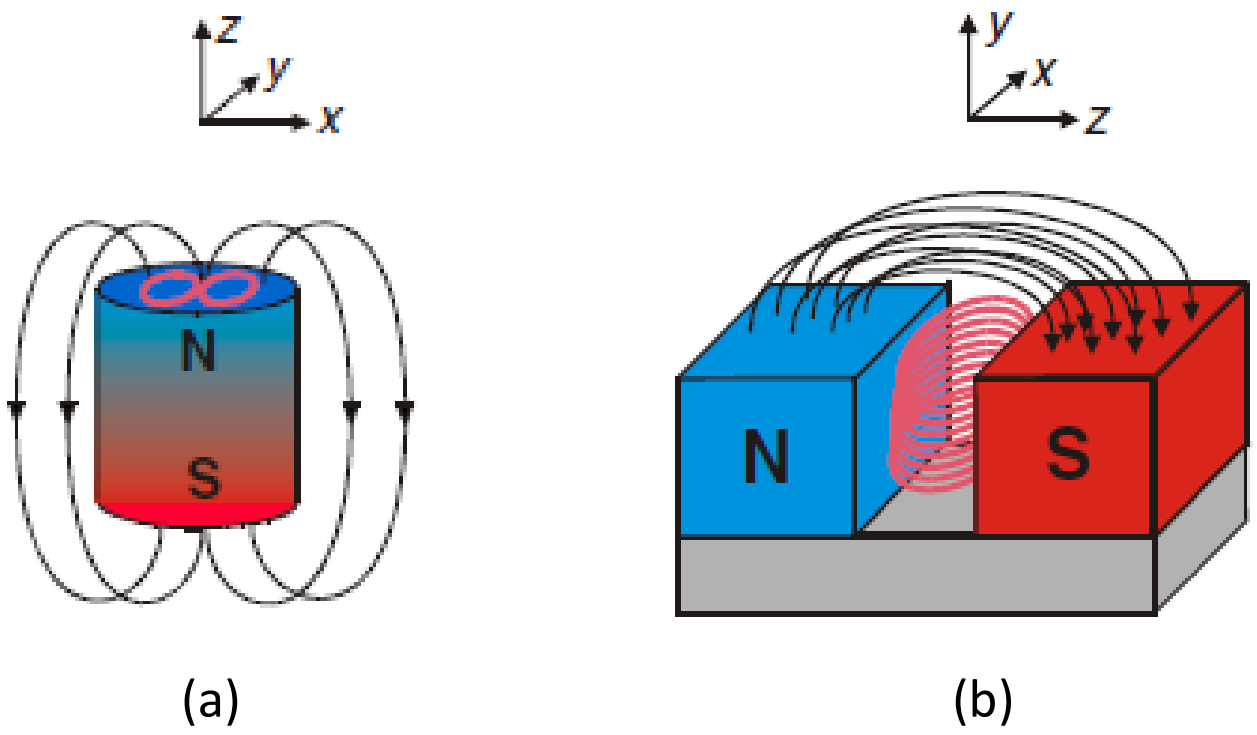

Fig. 13 Magnets for unilateral NMR (a) a simple bar magnet (b) a U-shaped open magnet

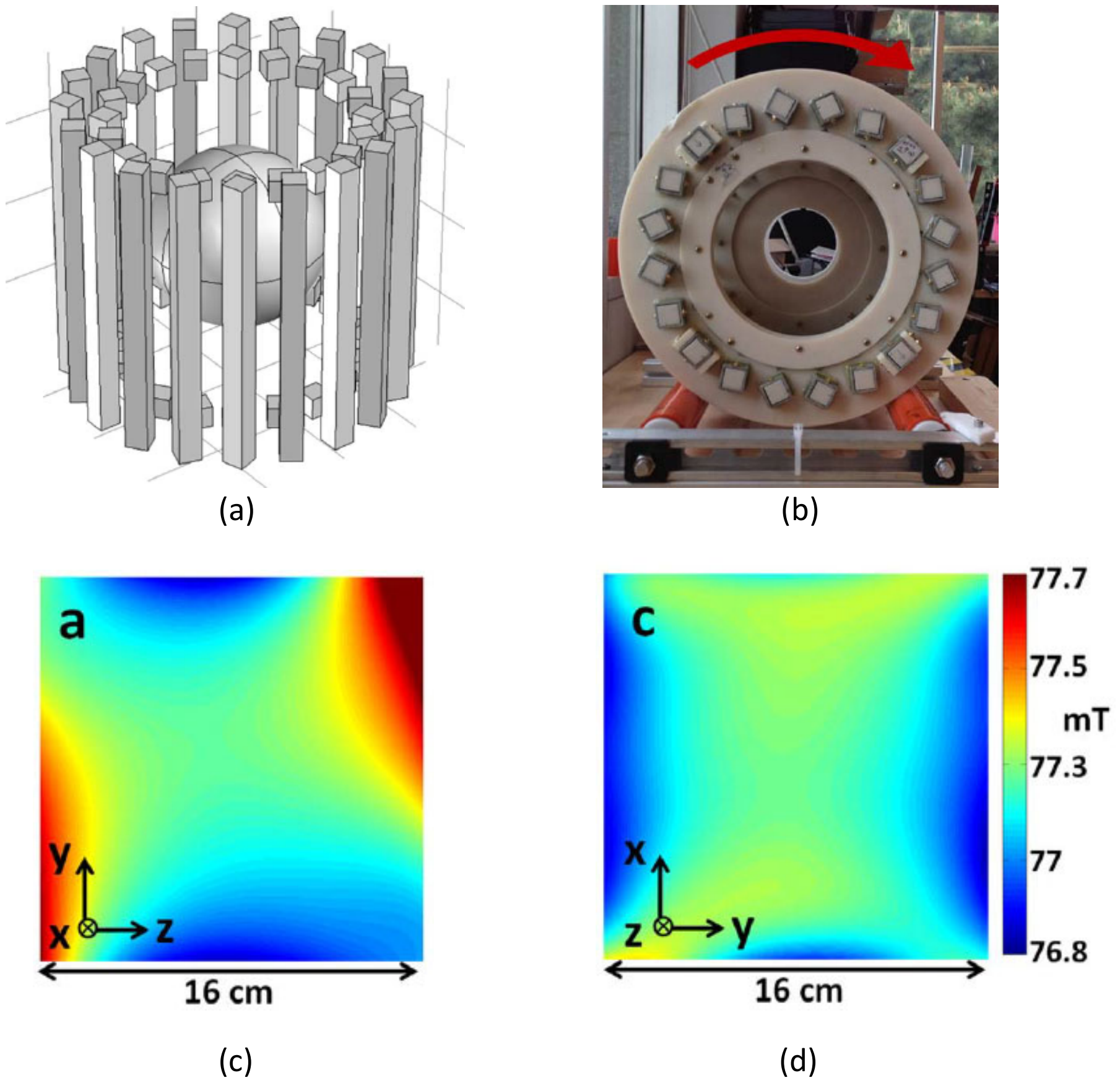

Fig. 14 A sparse dipolar Halbach cylinder for head imaging [20] (a) 3D view (b) side view (c) magnetic field distribution on the yz-plane (d) magnetic field distribution on the xy-plane

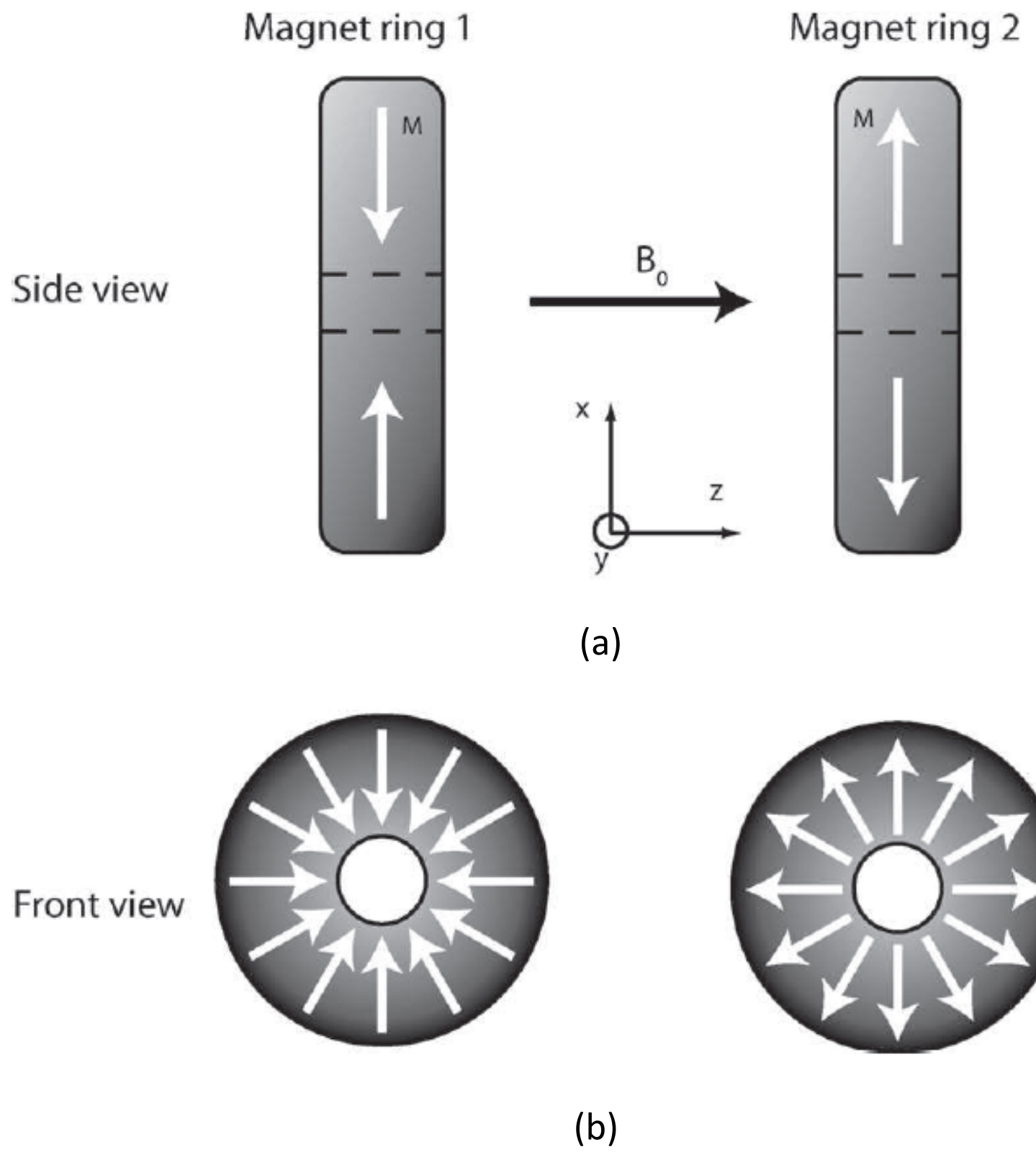

Fig. 15 Aubert ring pair [57]

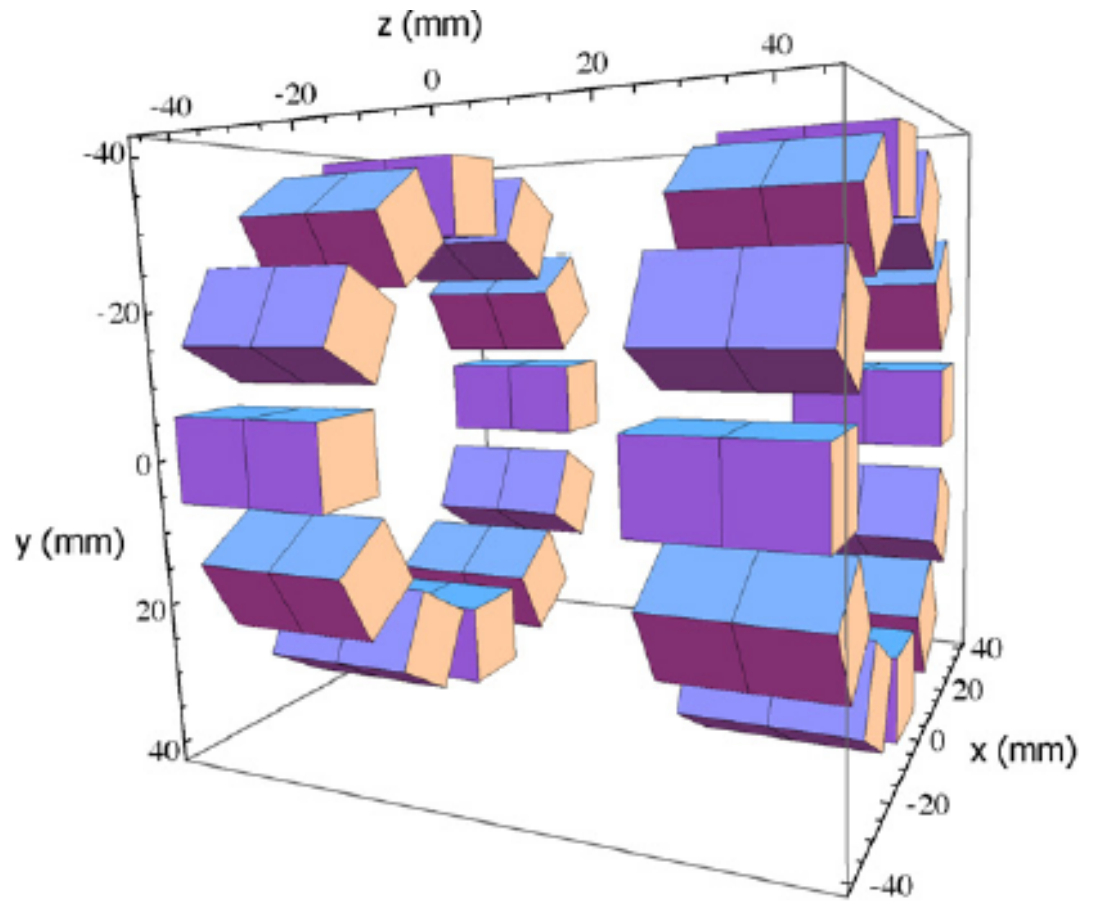

Fig. 16 A segmented Albert ring pair using magnet cubes [62]

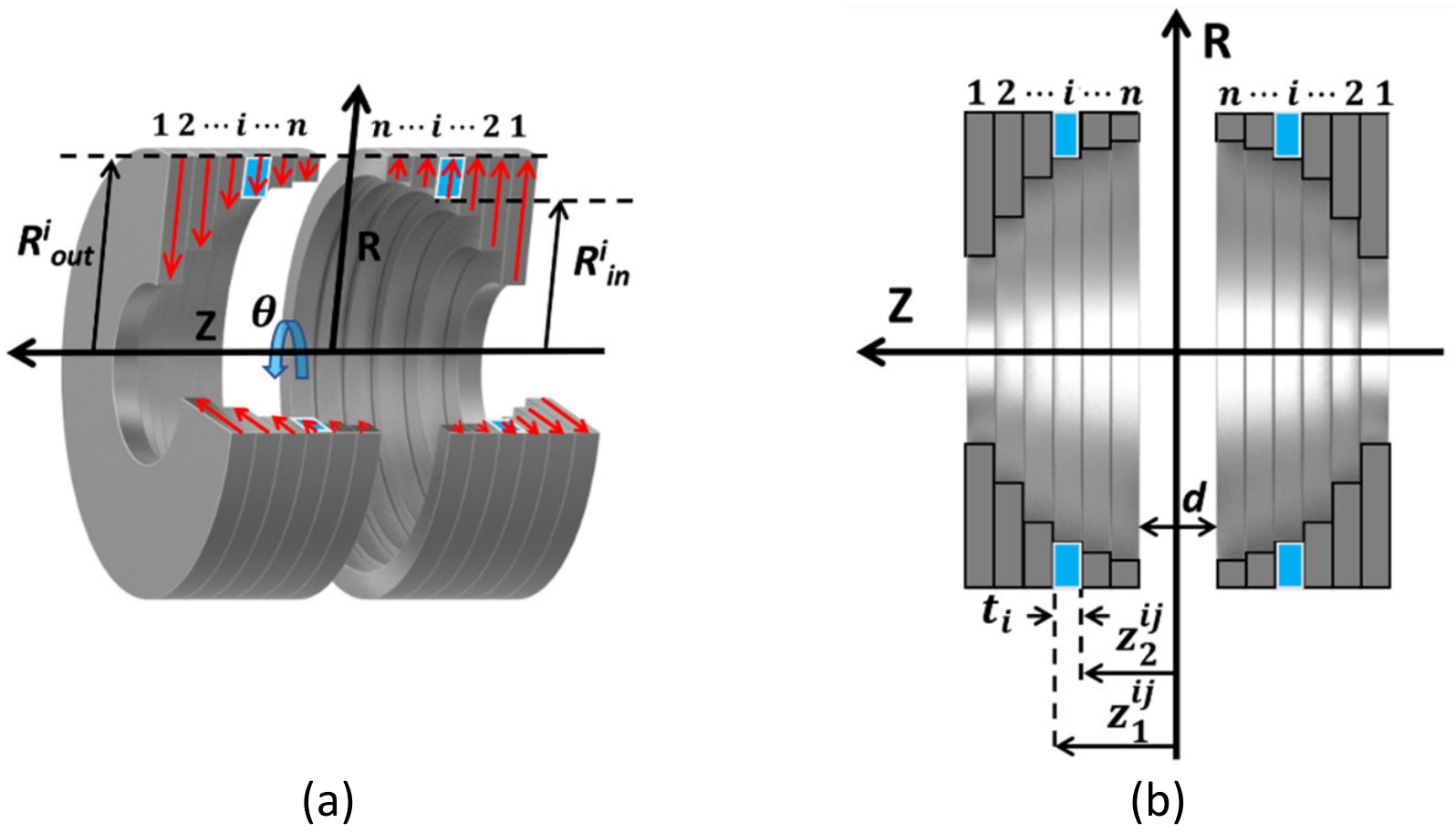

Fig. 17 Aubert Ring-pair aggregates for head imaging [65] (a) 3D view (b) side view

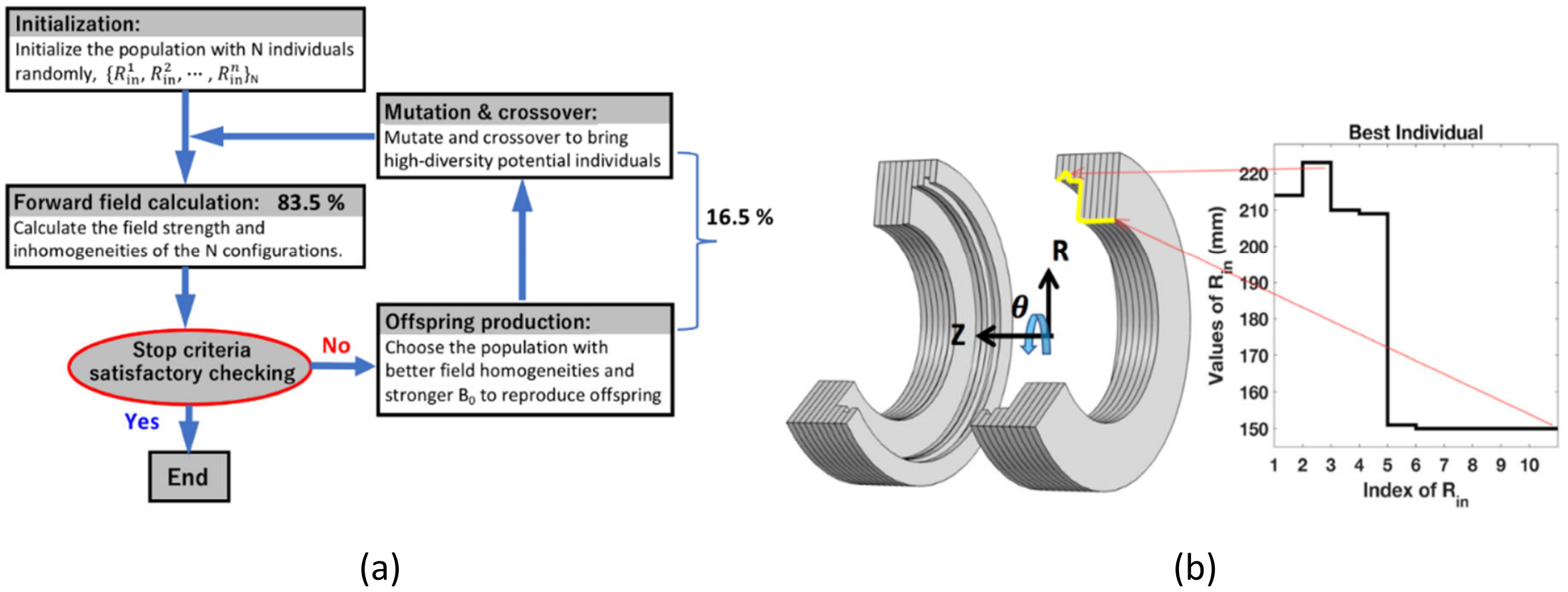

Fig. 18 Optimization using GA [65] (a) the GA flow (b) the results of optimization

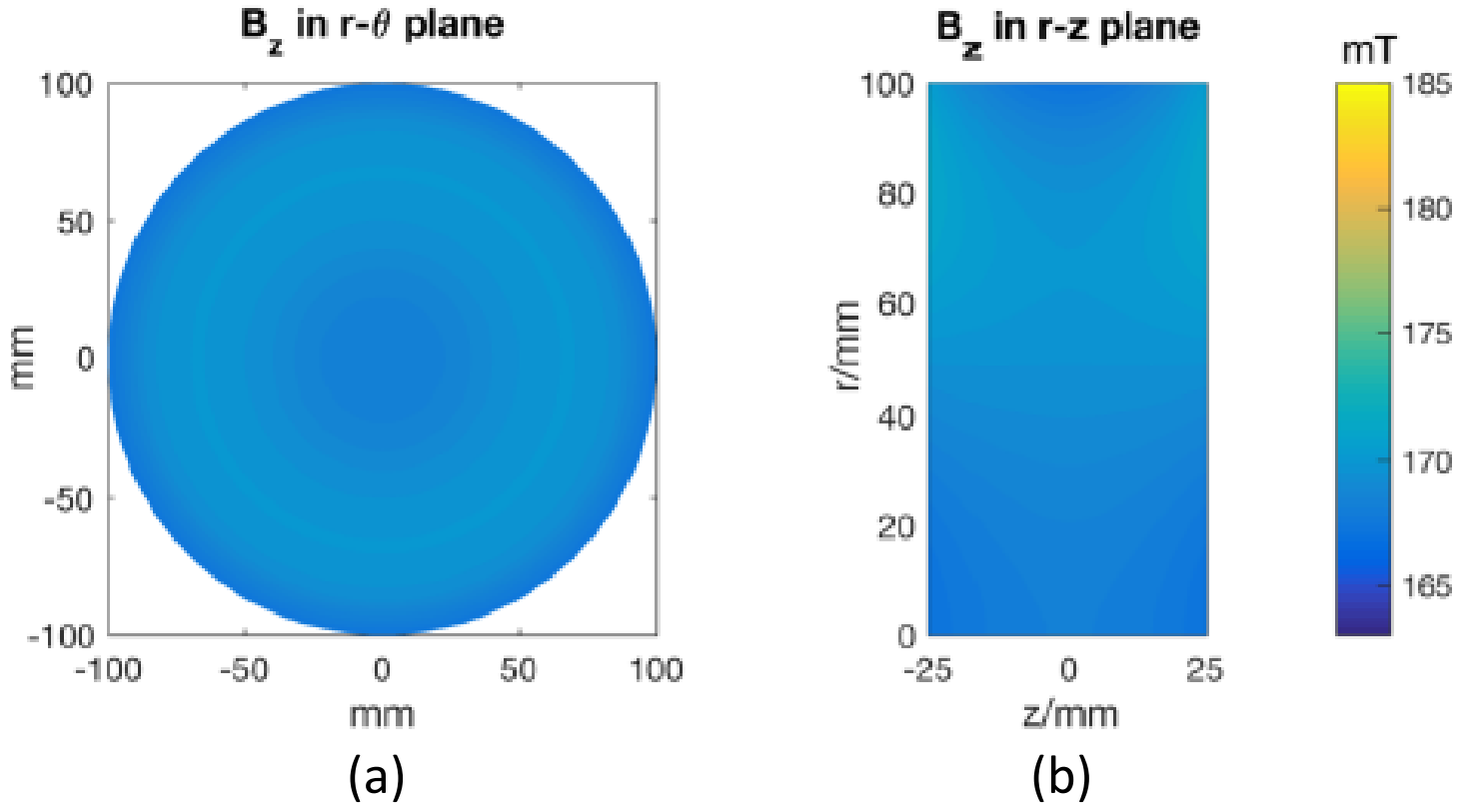

Fig. 19 The magnetic field generated by the proposed Aubert ring-pair aggregate in FoV in (a) the rϕ - plane, (b) the rz – plane [65].

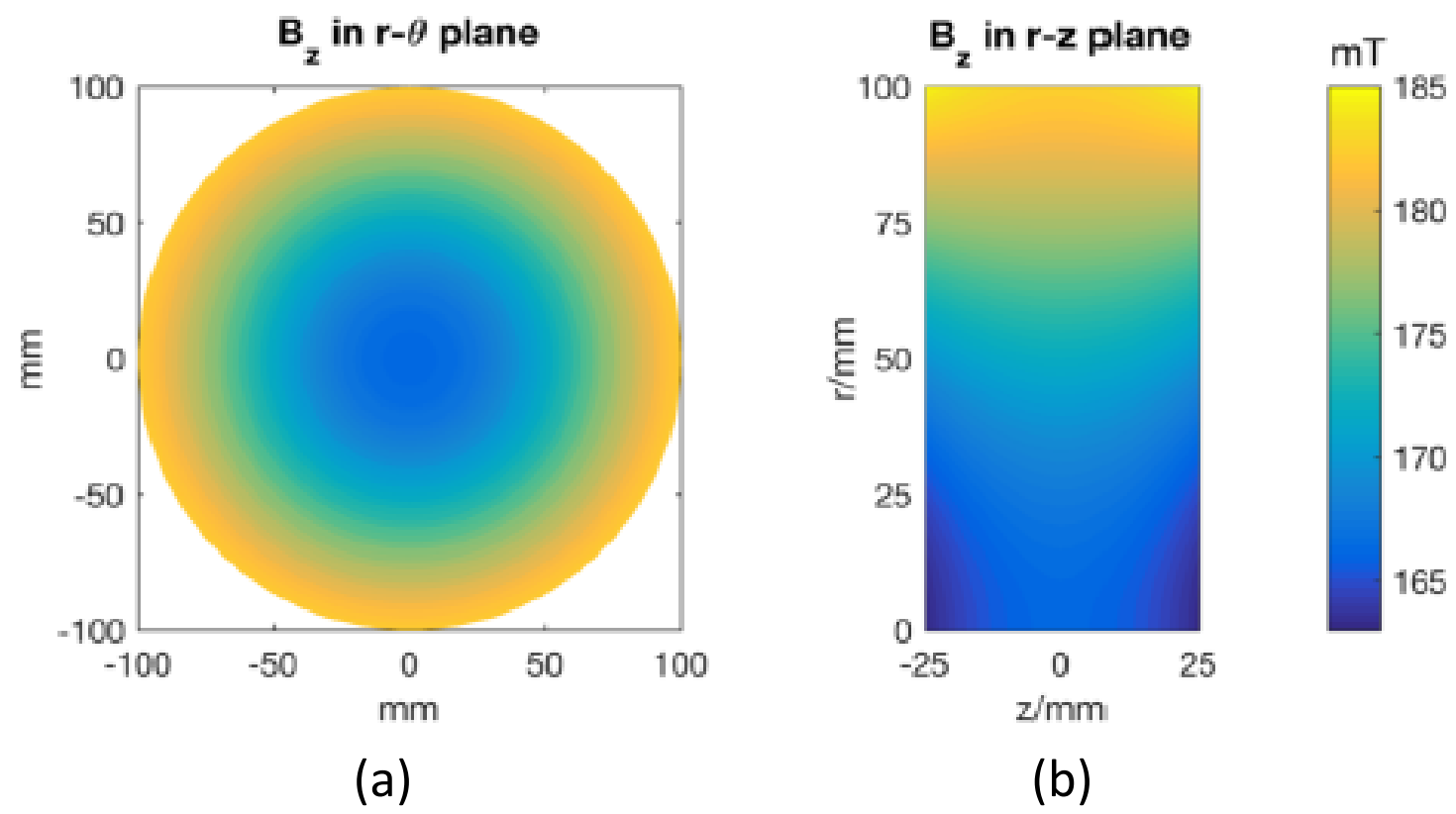

Fig. 20 The magnetic field generated by the original Aubert ring pair in FOV in (a) the rϕ - plane, (b) the rz – plane [65].

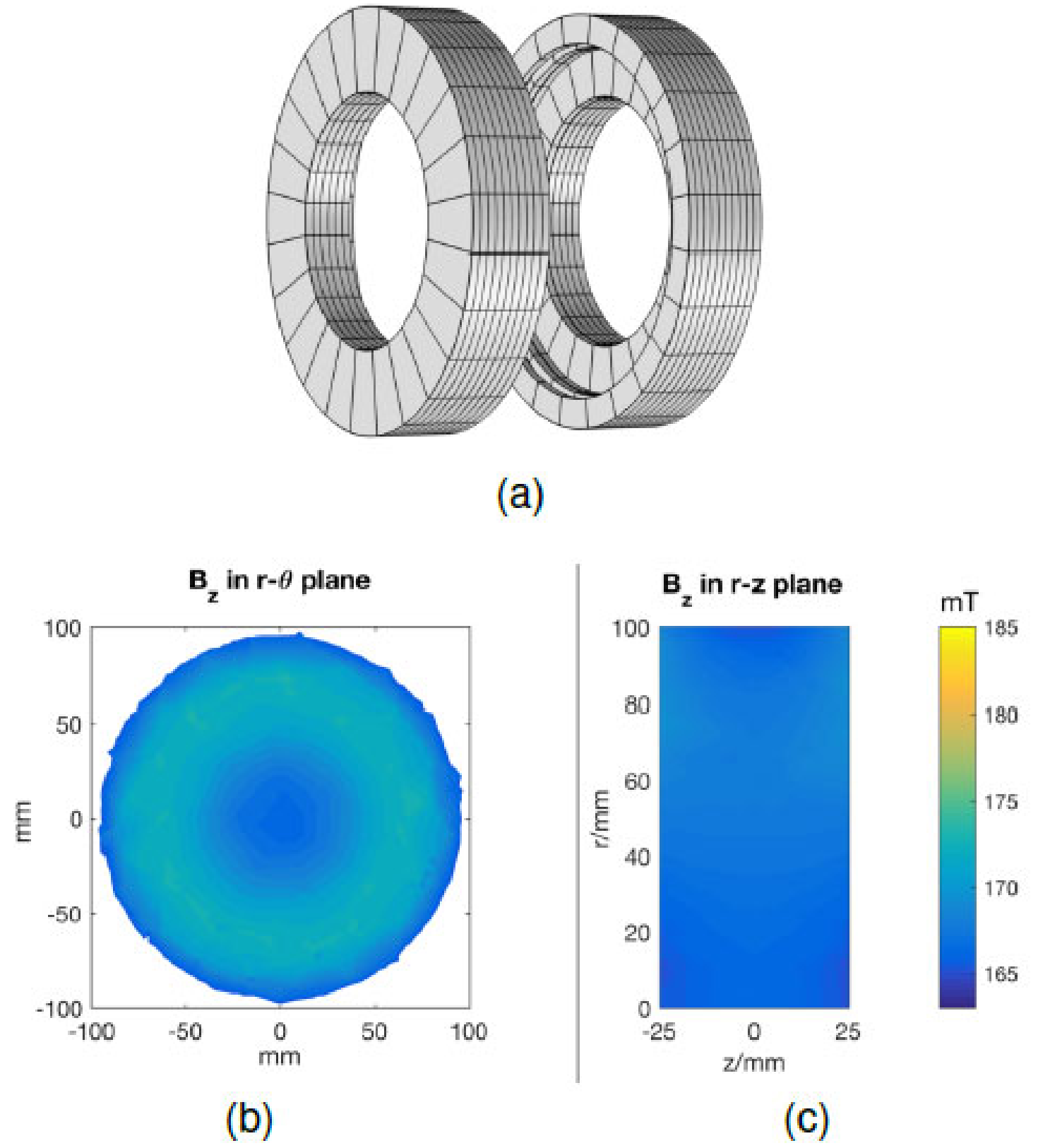

Fig. 21 The segmented optimized magnet array made up of fan-shaped magnets (a) The 3D view, the calculated magnetic field (b) on the $r\phi$-plane in FOV on the rz -plane in the FOV. COMSOL Multiphysics were used for the calculation [65].